\documentclass[amssymb,amsmath,aps,showpacs,floatfix,nofootinbib,10pt,onecolumn]{revtex4-1}
\usepackage{amssymb}
\usepackage{graphicx}
\usepackage{color}
\usepackage{latexsym}
\usepackage{epsfig}

\newcommand{\beqa}{\begin{eqnarray}}
\newcommand{\eeqa}{\end{eqnarray}}

\begin{document}

\voffset 1.25cm

\title{Tau neutrinos from ultracompact dark matter minihalos and constraints on the primordial curvature perturbations}

\author{Yupeng Yang$^{1,2}$,~Yan Qin$^{1}$}

\affiliation{$^{1}$ Collage of Physics and Electrical Engineering, Anyang Normal University, Anyang, 455000, China \\
$^{2}$ Joint Center for Particle, Nuclear Physics and Cosmology, Nanjing, 210093, China}

\date{\today}

\begin{abstract}
The observations and research on the neutrinos provide a kind of indirect way 
of revealing the properties of dark matter particles. For the detection of muon neutrinos, the main issue is 
the large atmospheric background, which is caused by the interactions between the cosmic rays and atoms 
within the atmosphere. Compared with muon neutrinos, 
tau neutrinos have a smaller atmospheric background 
especially for the downward-going direction. 
Except for the classical neutrino sources, 
dark matter particles can also annihilate into the neutrinos 
and are the potential high energy astrophysical sources. The annihilation 
rate of dark matter particles is proportional to the square of number density; 
therefore, the annihilation rate is large near the center of dark matter halos 
especially for the new kind of dark matter structures named 
ultracompact dark matter minihalos (UCMHs). In previous works, we have 
investigated the potential muon neutrino flux from UCMHs 
due to dark matter annihilation. Moreover, since the formation of UCMHs 
is related to the primordial density perturbations of small scales, 
we get the constraints on the amplitude of the primordial curvature perturbations 
of small scales, $1 \lesssim k \lesssim 10^{7}  ~\rm Mpc^{-1}$. 
In this work, 
we focus on the downward-going tau neutrinos from UCMHs due to 
dark matter annihilation. Compared with the background of tau neutrino flux we get the 
constraints on the mass fraction of UCMHs. 
Then using the limits on the mass fraction of UCMHs we 
got the constraints on the amplitude of the primordial 
curvature perturbations which are extended to the scale $k \sim 10^{8} ~ \rm Mpc^{-1}$ 
compared with previous results.

\end{abstract}


\maketitle


\section{Introduction}
As the main component of the Universe, dark matter has been confirmed 
by many observations while its nature remains unknown. 
At present, there are many dark matter models and the most researched model is the 
weakly interacting massive particles (WIMPs). 
According to the theory of WIMPs, they can annihilate into, e.g., 
photons ($\gamma$), electrons ($e^-$), positrons ($e^+$), and 
neutrinos ($\nu(\bar \nu)$)~\cite{darkmatter_1,darkmatter_2}. 
The observations and research on the particles produced 
by dark matter annihilation 
provide a way of indirectly detecting of dark matter particles. 
Moreover, the related research can be used 
to constrain the properties of dark matter particles, such as 
the mass ($m_{\rm DM}$) 
and thermally averaged cross section ($\left<\sigma v\right>$) of dark matter particles.  
For example, using the observations of $\gamma$-ray flux, it can be found that 
for the dark matter mass $m_{\rm DM} \lesssim 1\rm TeV$ 
the constraints on the 
thermally averaged cross section are $\left<\sigma v\right> \lesssim \rm 10^{-25} cm^2 s^{-1}$~\cite{dm_cons_1,dm_cons_2,dm_cons_3,liuwei}. 
For large dark matter mass, e.g., $m_{\rm DM} \gtrsim 1\rm TeV$, 
the constraints on $\left<\sigma v\right>$ from the $\gamma$-ray observations are weaker than that from 
the neutrino observations~\cite{dm_cons_neutrino_1,dm_cons_neutrino_2,Adrian-Martinez:2015wey}. 
The main way of using the neutrinos to research the properties of dark matter particle is 
through the observations and studies on muon neutrinos ($\nu_{\mu}$)~\cite{dm_cons_neutrino_2,Bottino:2004qi,Bottino:1994xp,Blennow:2013pya}. 
The main flaw of observations on $\nu_{\mu}$ 
is the large atmospheric background, which is 
caused by the interactions between the cosmic rays and atoms 
within atmosphere~\cite{tau-downward,dm_decay_neutrino}. For electron neutrinos, $\nu_e$, the 
atmospheric background is also large and the cascade effects make the detection 
of $\nu_e$ very difficult~\cite{Kumar:2011hi,Guo:2015hsy}. Compared with the muon and electron neutrino 
the atmospheric background of the tau neutrino ($\nu_{\tau}$) is small especially in 
the direction of $\rm cos\theta_{Z} > 0$, where $\theta_{Z}$ is the zenith angle. In particular, 
for $\rm cos\theta_{Z} \gtrsim 0.5$, 
the atmospheric tau neutrino fluxes are 3 orders smaller than 
the atmospheric muon and electron neutrino background~\cite{tau-downward}. 
For $\rm cos\theta_{Z} \gtrsim 0.7$, the atmospheric tau neutrino fluxes are even smaller than 
that from solar corona interaction and galactic neutrino flux~\cite{tau-downward}. Although the cascade effects 
of the tau neutrino make detection difficult, with the development of detection 
and a statistical method the detection of the tau neutrino has become possible~\cite{Xu:2017yxo,Conrad:2010mh,Cuoco:2006qd}. 

Dark matter plays an important role in the process of the structure formation of the Universe. 
It is well known that the structures of the Universe are the evolutionary results of the early 
density perturbations with an amplitude $\delta {\rho}/{\rho} \sim 10^{-5}$~\cite{cmb_2}. If the amplitude of early 
density perturbations is larger than $\sim$ 0.3, the primordial black 
holes (PBHs) can be formed~\cite{pbh}. Recently, 
the authors of~\cite{ucmhs_1} suggested that ultracompact dark matter minihalos can be 
formed in the early time if the amplitude of primordial density perturbations is in the range of 
$\delta {\rho}/{\rho} \sim 10^{-3} - 0.3$. 
After the formation of ultracompact dark matter minihalos (UCMHs), during the radiation dominated epoch, 
the mass of UCMHs keeps unchanged nearly due to the Meszaros effect. 
After the redshift of equality of radiation and matter, the mass of UCMHs scales as 
$M_{\rm UCMHs} \sim 1/(1+z)$. For the density profile 
of UCMHs, one dimension simulation shows that it scales as 
$\rho_{\rm UCMHs}(r) \sim r^{-9/4}$~\cite{ucmhs_1}. Compared with the 
mostly used dark matter halo models, such as the Navarro-Frenk-White(NFW) model~\cite{NFW}, 
the density profile of UCMHs is steeper than that of the NFW profile especially for $r \to 0$, 
$\rho_{\rm NFW}(r)\sim r^{-1}$. The annihilation rate of dark matter particles is 
proportional to the square number density; therefore, it is excepted that 
the annihilation rate of dark matter particles is larger within UCMHs 
than that within the classical dark matter halos. In Ref.~\cite{scott_prl}, the authors 
investigated the $\gamma$-ray flux from UCMHs due to dark matter annihilation. 
They found that the $\gamma$-ray flux from UCMHs formed during the $e^{+}e^{-}$ 
phase transition can excess the threshold of Fermi or EGRET observations 
for some dark matter annihilation channels. 
Besides the $\gamma$-ray flux, in theory, 
the neutrinos can also be emitted from UCMHs due to dark matter annihilation 
especially for the lepton channels. In Ref.~\cite{yyp_neutrino}, the authors found that the muon neutrino flux from UCMHs 
can exceed the atmospheric muon neutrino flux. With no 
detection of excess of $\gamma$-ray flux the upper constraints on the 
abundance of UCMHs are 
obtained, $f_{\rm UCMHs} < 10^{-7}$~\cite{Bringmann_1}. Similar to the $\gamma$-ray flux, 
the research on neutrino flux 
can also be used to do the studies on the abundance of UCMHs~\cite{yyp_neutrino,yyp_ijmpa}. 
Besides the research on the particles produced by dark matter 
annihilation, in Refs.~\cite{fangdali,ucmhs_pulsar}, the authors investigated the gravitational effects 
caused by UCMHs and got the constraints on the abundance of UCMHs. 

The formation of UCMHs is related to the primordial 
perturbations. After obtaining the limits on the abundance 
of UCMHs, one can then use the limits on the abundance of UCMHs 
to get the constraints on the primordial curvature perturbations~\cite{Bringmann_1,yyp_neutrino,yyp_ijmpa,fangdali}. 
It is well known that the structure formation is related 
to the primordial curvature perturbations, 
$\mathcal{P}_{\mathcal{R}}(k)$, which stand for the amplitude of the 
primordial curvature perturbations. At present, the constraints on  
$\mathcal{P}_{\mathcal{R}}(k)$ are mainly on large scales($k \sim 10^{-4} - 1\ \rm Mpc^{-1}$) 
and from the observations and research 
on the CMB, Lyman-$\alpha$ forest and, large scale structure~\cite{cmb_2,lyman,large}. 
All of these observations 
show a nearly scale-invariant spectrum of primordial 
perturbations with $\mathcal{P}_{\mathcal{R}}(k) \sim 10^{-9}$, which is predicted 
by the popular inflation theory. On small scales, $k \sim 1-10^{20}
\ \rm Mpc^{-1}$, the constraints on $\mathcal{P}_{\mathcal{R}}(k)$ 
are mainly from the research on PBHs, $\mathcal{P}_{\mathcal{R}}(k) \lesssim 
10^{-2}$~\cite{Josan:2009,Bringmann_1}. Similar to PBHs, the UCMHs can be formed in very early times; 
therefore, through the research on the UCMHs one can get the constraints on the primordial curvature 
perturbations of small scales. In Ref.~\cite{Bringmann_1}, with no detection of $\gamma$-ray flux from UCMHs, 
the authors got the constraints 
on $\mathcal{P}_{\mathcal{R}}(k)$ on scales 
$k \sim 5-10^{8}\ \rm Mpc^{-1}$, $\mathcal{P}_{\mathcal{R}}(k)\lesssim 10^{-6}$. 
Those constraints are better than 
that of PBHs. Similar to the $\gamma$-ray flux, in previous work, 
we got the comparable results through investigating the potential 
muon neutrino flux from UCMHs due to dark matter annihilation~\cite{yyp_neutrino}. 
According to the theory, $\nu_{e}$ and $\nu_{\tau}$ can also be produced 
in the process of dark matter annihilation. Moreover, the oscillation 
property of neutrinos can also result in the conversion among three 
flavors of neutrino. As mentioned above, the downward-going tau neutrinos can also 
be used to look for the neutrino signals from dark matter 
annihilation due to the lower atmospheric background. In this paper, 
we investigate the potential tau neutrino flux from UCMHs 
due to dark matter annihilation and focus on the downward-going 
tau neutrino flux($\rm cos\theta_{Z} >0$). By comparing with the 
atmospheric $\nu_{\tau}$ background, we obtained the potential constraints 
on the abundance of UCMHs for the IceCube experiment. 
Then using the limits on the abundance of UCMHs we get the constraints 
on the primordial curvature perturbations of small scales. 

This paper is organized as follows. The tau neutrino 
background is reviewed in Sec. II. In Sec. III, we discuss 
the main properties of UCMHs and the potential tau neutrino flux from them. 
In Sec. IV, through comparing with the background, the potential 
constraints on the abundance of UCMHs are obtained and using 
these constraints we then get the upper limits on the primordial 
curvature perturbations of small scales. The conclusions and discussions 
are presented in Sec. IV.

\section{Background of tau neutrino flux}
There are several sources for the background of tau neutrinos. The main one 
is the atmospheric background and it is mainly due to the oscillation 
of the muon neutrinos. This background is 
lower than that of the electron and muon neutrinos especially for 
$\rm cos\theta_{Z} >0$. For example, for $\rm cos\theta_{Z} \gtrsim 0.5$, 
the background of $\nu_{\tau}$ is about 3 orders lower than 
that of $\nu_{\mu}$ or $\nu_e$~\cite{tau-downward}. For this background, 
we use the form given in Ref.~\cite{atm_1,atm_2} as 

\beqa
\frac{d\Phi_{\nu_\mu}}{dE_{\nu_\mu}d\Omega}=N_{0}E^{-\gamma -1}_{\nu_\mu}
\left(\frac{a}{1+bE_{\nu_\mu}cos\theta}+\frac{c}{1+eE_{\nu_\mu}cos\theta}\right)
\mathrm{GeV^{-1} km^{-2} yr^{-1} sr^{-1}},
\eeqa
where $\theta$ is the zenith angle, 
$N_{0}=1.95\times 10^{17}(1.35\times 10^{17})$ for $\nu_\mu(\bar \nu_\mu)$, 
$\gamma =1.74, a=0.018, b=0.024 \mathrm{GeV^{-1}}, c=0.0069, 
e=0.00139 \mathrm{GeV^{-1}}$. The conversion probability of $\nu_\mu$ 
into $\nu_{\tau}$ can be written as

\beqa
P(\nu_{{\mu}\to{\tau}})=\rm sin^2 2\theta_{atm}sin^2\left(1.27
\frac{\Delta m^2_{atm}L}{E_{\nu}}\right),
\eeqa
where L is the propagation length of neutrinos after being produced 
in the atmosphere~\cite{dm_decay_neutrino}. 
For the parameters related to the neutrino oscillations, following Refs.
~\cite{tau-downward,dm_decay_neutrino}, 
we have set $\rm sin^2 2\theta_{atm}=1$, $\lvert \rm \Delta m^2_{atm}\rvert 
= 2.4 \times 10^{-3} eV^2$. 

In addition to the $\nu_{\tau}$ flux coming from the conversion of 
$\nu_{\mu}$, the decay of charmed 
particles produced in the atmosphere provides another background of $\nu_{\tau}$ and 
this can be parametrized as~\cite{tau_origi,tau-downward,dm_decay_neutrino}

\beqa
\rm log_{10}\left[E^{3}_{\nu}\frac{d\phi_{\nu}}{dE_{\nu}}/
\left(\frac{GeV^2}{cm^2 \ s \ sr}\right)\right]=-A+B \mathnormal{x}
-C\mathnormal{x}^2 - D\mathnormal{x}^3,
\eeqa
where $\mathnormal{x}=\rm log_{10}(E_{\nu}[GeV])$, A=6.69, B=1.05, C=0.150 and 
D=-0.00820. 

Neutrinos can also be produced in the solar corona by cosmic-ray 
collisions. This neutrino flux has been studied in Ref.~\cite{tau_solar}, the 
$\nu_{e}$ and $\nu_{\mu}$ flux can be written as 

\beqa
\frac{d\phi_{\nu}}{dE_{\nu}}=\rm N_{0}\frac{\left(E_{\nu}[GeV]\right)^
{-\gamma-1}}{1+A\left(E_{\nu}[GeV]\right)}\left(GeV\ cm^2\ s\right)^{-1},
\eeqa
which is valid for $\rm 10^{2}GeV \leq E \leq 10^{6}GeV$. The numerical 
values of the coefficients $N_0$, $A$ and $\gamma$ can be found in Ref.~\cite{tau_solar}. 
Recently, the authors of Refs.~\cite{solar_cor_update_1,solar_cor_update_2,solar_cor_update_3} 
have revisited this neutrino flux and updated the results. In this paper, we have used 
these new results for our calculations.

In Ref.~\cite{tau_galactic}, the authors discussed that the tau neutrinos can also originate 
from a galactic plan. Considering the oscillations of neutrinos the tau neutrino flux 
can be parametrized as

\beqa
\frac{d\phi_{\nu_\tau}}{dE}=9\times 10^{-6}\rm \left(GeV\ cm^2\ s\ sr\right)^{-1}
\left(E[GeV]\right)^{-2.64}
\eeqa
which is valid in the energy range $\rm 1\ GeV \leq E \leq 10^{3}\ GeV$. 

For the background of tau neutrino flux, the main component is the conversion 
of atmospheric muon neutrino. In this paper, we considered the total flux mentioned above 
for our calculations. Moreover, we considered downward-going tau neutrino flux as $\rm cos\theta > 0$.

\section{formation of UCMHs and tau neutrino flux from UCMHs due to dark matter 
annihilation}

The UCMHs can be formed in the early Universe, e.g., $z\sim 1000$, 
if the primordial density perturbations are in the range of $10^{-3}< \delta \rho/\rho<0.3$~\cite{ucmhs_1,ucmhs_2}. 
After formation the mass of UCMH changes as~\cite{ucmhs_1}

\beqa
M_{\rm UCMH}(z)=M_{i}\frac{1+z_{eq}}{1+z},
\eeqa
where $M_i$ is the initial mass within the perturbation scale. 
The results of one-dimension simulation show that the density profile of the UCMH is in the following form
~\cite{ucmhs_1,scott_prl,Bringmann_1}
\footnote{Recently, the authors of Ref.~\cite{ucmhs_2017_profile} have done the 3D simulations and found that 
the NFW profile is a better fit to the density profile of the UCMH.},

\beqa
\rho(r,z)=\frac{3f_{\chi}M_{\rm UCMH}(z)}{16\pi R(z)^{3/4}r^{9/4}},
\eeqa
where $f_{\chi}=\frac{\Omega_{\rm DM}}{\Omega_{\rm b}+\Omega_{\rm DM}}=0.83$~\cite{Komatsu:2008hk}, 
$R(z)$ is the radius of UCMH, 

\beqa
R(z)=0.019 \left(\frac{1000}{1+z}\right)\left(\frac{M_{\rm UCMHs}(z)}{M_{\odot}}\right)^{1/3}\ \rm pc. 
\eeqa
After the redshift, e.g., $z \sim 10$, the structure formation is 
dominated in the Universe. Therefore, we set $z_{\rm stop}=10$ and at that time 
the mass of UCMHs stops increasing~\cite{ucmhs_1,scott_prl,Bringmann_1,yyp_neutrino,fangdali,scott_2015,ucmhs_pulsar}. 
The dark matter annihilation rate is proportional to 
the square of number density; therefore, the inner density profile of UCMH is very important 
for the related studies~\cite{ucmhs_inner}. Generally, one treats the density of UCMH
as a constant value for radius $r\lesssim r_{\min}$, 
$\rho_{\rm UCMH}(r\lesssim r_{\rm min})= cons.$~\cite{scott_prl,Bringmann_1,yyp_neutrino}. 
Here we considered two factors that have remarkable effects on $r_{\rm min}$. 
One factor is to consider the conservation of angular momentum of 
dark matter particles. After the formation of UCMHs, dark matter particles 
accrete 
on UCMHs by radial infall. Considering the conservation of angular momentum, 
the cutoff radius $r_{\rm min}$ can be written as~\cite{Bringmann_1} 

\beqa
r_{\rm min}= 5.1\times 10^{-7} \mathrm{pc} \left(\frac{1000}{1+z}\right)^{2.43}
\left(\frac{M^{0}_{\rm UCMH}}{M_{\odot}}\right)^{0.27}.
\eeqa

Another factor that can affect the center density of UCMH is the 
annihilation of dark matter particles. 
For this factor, following Refs.~\cite{yyp_neutrino,scott_prl,Bringmann_1}, we truncate the radius at $r_{\rm cut}$. 
For $r < r_{\rm cut}$, the density profile of UCMHs is 

\beqa
\rho_{\rm UCMHs}(r<r_{\rm cut})=\frac{m_{\rm DM}}{\left<\sigma v\right>(t-t_i)},
\label{rho_cen}
\eeqa
where $m_{\rm DM}$ and $\left<\sigma v\right>$ are the 
mass and thermally averaged cross section of the dark matter particle, respectively. 
$t$ is the cosmic time and $t_i$ is the formation time of UCMHs. For the parameters considered in this work, 
we find $r_{\rm cut} \gtrsim r_{\rm min}$, therefore, we adopt $r_{\rm cut}$ for our 
calculations. More detailed discussions about the center density profile can be found in, e.g., Refs.~\cite{Bringmann_1,profile}.

\begin{figure}
\epsfig{file=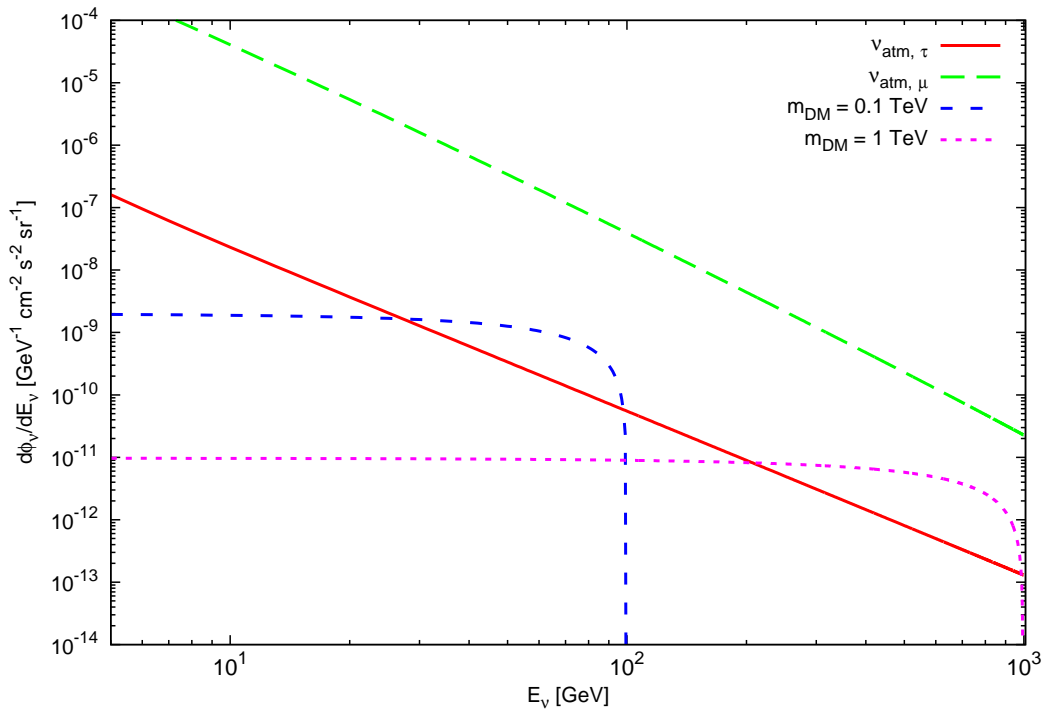,width=0.8\textwidth}
\caption{Tau neutrino flux from the UCMH due to dark matter annihilation for 
the $\tau^{+}\tau^{-}$ channel. We have set the dark matter mass 
$m_{\rm DM}$=0.1 (blue short-dashed line) and 1 TeV (purple dot line), and 
the thermally averaged cross section $\left<\sigma v\right>=3 \times 10^{-26} \ \rm cm^{3} \ s^{-1}$. 
In this plot, we have considered the UCMH formed during the phase transition named 
$e^{+}e^{-}$ annihilation. The distance of the UCMH is $d_{\rm UCMH}$ = 0.1 kpc. 
For comparison, in addition to the $\nu_{\tau}$ background flux (red solid line) 
the $\nu_{\mu}$ background flux is also shown (green long-dashed line).}
\label{diffflux}
\end{figure}

The neutrino flux from the UCMH due to dark matter annihilation can be written as~\cite{yyp_neutrino}

\beqa
\frac{d\phi_{\nu}}{dE_{\nu}d\Omega}=\frac{1}{8\pi}\frac{dN_{\nu}}{dE_{\nu}}
\frac{\left<\sigma v\right>}{m^2_{\rm DM}}\frac{1}{d^2_{\rm UCMH}}\int^{r_{\rm max}}_{r_{\rm min}} 
\rho^2_{\rm UCMH}(r,z_{\rm stop})4\pi r^2dr,
\label{dphi}
\eeqa
where $d_{\rm UCMH}$ is the distance of the UCMH from the Earth, 
$dN_{\nu}/dE_{\nu}$ is the neutrino number per dark matter annihilation 
and can be obtained from the public code DarkSUSY\footnote{http://www.darksusy.org/}. 
The tau neutrino 
flux from the UCMH is shown in Fig.~\ref{diffflux}. In this figure, we considered 
the $\tau^{+}\tau^{-}$ 
annihilation channel and set the dark matter mass $m_{\rm DM}$=0.1 (blue short-dashed line) and 1 TeV 
(purple dot line), the thermally averaged cross section $\left<\sigma v\right>=3 \times 10^{-26} \ \rm cm^{3} \ s^{-1}$. 
In this plot, we considered the UCMH formed during the phase transition named 
$e^{+}e^{-}$ annihilation and the distance is $d_{\rm UCMH}$ = 0.1 kpc. 
In addition to the background of the $\nu_{\tau}$ flux, for comparison, the backgrounds of the $\nu_{\mu}$ flux are also shown. 
As shown in Fig.~\ref{diffflux}, the $\nu_{\tau}$ flux from the UCMH due to dark matter annihilation is higher 
than the $\nu_{\tau}$ background but lower than $\nu_{\mu}$ background. 

\section{constraints on the fraction of UCMHs and primordial curvature perturbations}

Detecting and researching the neutrinos provide an important way of indirectly searching for the dark matter 
particles~\cite{Bergstrom:1996kp,Arcadi:2017kky}. At present, a typical 
way is to detect the muon neutrinos, especially for the upward-going 
neutrino flux. The detection of electron neutrinos is also considered, see e.g.,~\cite{Kumar:2011hi}. 
For the background, the muon neutrino is dominated and the background of electron neutrino 
is in the same level compared with that of the muon neutrino~\cite{tau-downward}. 
Compared with the muon and electron neutrino, the background of tau neutrino flux is lower 
especially in the direction of $\rm cos\theta_{Z} \gtrsim 0.5$. 
The main interaction for $\nu_{\tau}$ is the charged-current interaction and 
for the IceCube or ANTARES experiment the cascade events for 
the detection of $\nu_{\tau}$ can be written as~\cite{tau_detec}

\beqa
N_{\nu_\tau}=\int d\Omega \int_{E_{\rm min}}^{E_{\rm max}} dE \rho
N_{A}V_{\rm eff}\left(\sigma_{\nu N}(E)_{\rm CC}\frac{d\phi_{\nu}}{dE_{\nu}d\Omega}\right),
\eeqa
where $V_{\rm eff}$ is the effective volume of the detection\cite{2009arXiv0907.2263W,Coyle:2017pnl}. 
$\rho$ is the density of ice for IceCube and water for ANTARES, $N_{A} = 6.022 \times 
10^{23}$ is Avogadro's number. $\sigma_{\nu N}(E)_{\rm CC}$ is the charged-current cross section and we adopt the form given in Ref.~\cite{sigma_cc}. 
For the IceCube and ANTARES experiments, 
we set the energy bin as $\left [\mathrm{max}\left (E_{\rm thresh}, m_{\rm DM}/5\right ), 
m_{\rm DM}\right ]$ for the events calculations.

Compared with the background, 
$2\sigma$ statistical significance can be obtained as~\cite{tau-downward,Zhu:2008yh}

\beqa
\zeta \equiv \frac{N_{S}}{\sqrt{N_{S}+N_{B}}},
\eeqa
where $N_{S}$ and $N_{B}$ are the neutrino events from UCMH and background respectively. 

\begin{figure}
\epsfig{file=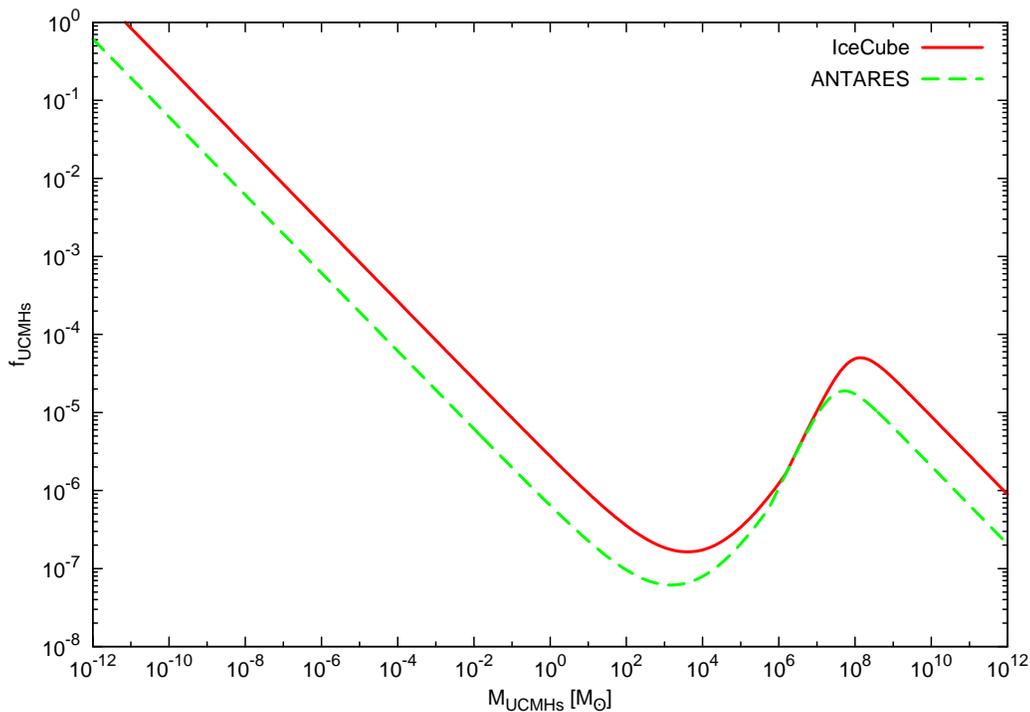,width=0.8\textwidth}
\caption{Upper limits (95\% C.L.) on the mass fraction of UCMHs 
for the downward-going tau neutrino flux for IceCube (red solid line) 
and ANTARES (green dashed line). 
Here we have set the dark matter 
mass $m_{\rm DM} = 1 \rm TeV$ and the thermally averaged cross section 
$\left<\sigma v\right>=3 \times 10^{-26} \ \rm cm^{3} \ s^{-1}$. 
The annihilation channel is $\tau^{+}\tau^{-}$.}
\label{fraction_plot}
\end{figure}

The fraction of UCMHs can be calculated using the following formula~\cite{Bringmann_1}

\beqa
f_{\rm UCMHs} = \frac{f_{\chi}M_{\rm UCMH}^{0}}{M_{\rm MW}}\frac{\rm 
log(1-\frac{\mathnormal y}{\mathnormal x})}
{\rm log(1-\frac{M_{{\mathnormal d}<{\mathnormal d}_{\rm obs}}}{M_{\rm MW}})},
\label{fraction}
\eeqa
where $M_{r<d_{\rm obs}}$ is the mass within the radius $d_{\rm obs}$ 
which is the distance on which the neutrino signals from UCMH would be 
observed by the detector\footnote{If the distance of UCMH is larger than the radius 
of Milky Way(MW), the mass within $d_{\rm obs}$ is written as~\cite{Josan:2009,yyp_neutrino} 
$M_{d<d_{\rm obs}}=\frac{4\pi}{3}(d_{\rm obs}^{3}-d_{max,MW}^{3})\rho_{\rm DM}
+M_{\rm MW}$, where $M_{\rm MW}$ is the mass of MW.}. 
In this work, 
we use the NFW profile for the dark matter halo model of the Milky Way. 
Using the above equations, one can obtain the values 
of $d_{\rm obs}$ for $2\sigma$ statistical significance for a different 
mass of the UCMH. Then the limits on the fraction of UCMHs can be obtained 
using Eq.~(\ref{fraction}) and the results are shown in Fig.~\ref{fraction_plot}. 
These constraints are comparable to the previous results that are obtained using the 
gamma-ray flux, e.g. Ref.~\cite{Bringmann_1}. The background of $\nu_{\tau}$ flux is 
lower than that of $\nu_{\mu}$ flux, therefore, compared with previous works, the constraints are extended to 
the smaller mass\footnote{Considering the effect of kinetic decoupling 
of WIMP, there is the smallest mass of UCMH~\cite{Bringmann_1}.}, 
$M_{\rm UCMH} \sim 10^{-11} M_{\odot}$. 
Similar to the constraints on the basic parameters ofthe  dark matter particle~\cite{Adrian-Martinez:2015wey}, 
for ANTARES, the constraints on the fraction of UCMHs are 
about 4 factors better than that of IceCube 
for the most mass ranges of UCMHs.

\begin{figure}
\epsfig{file=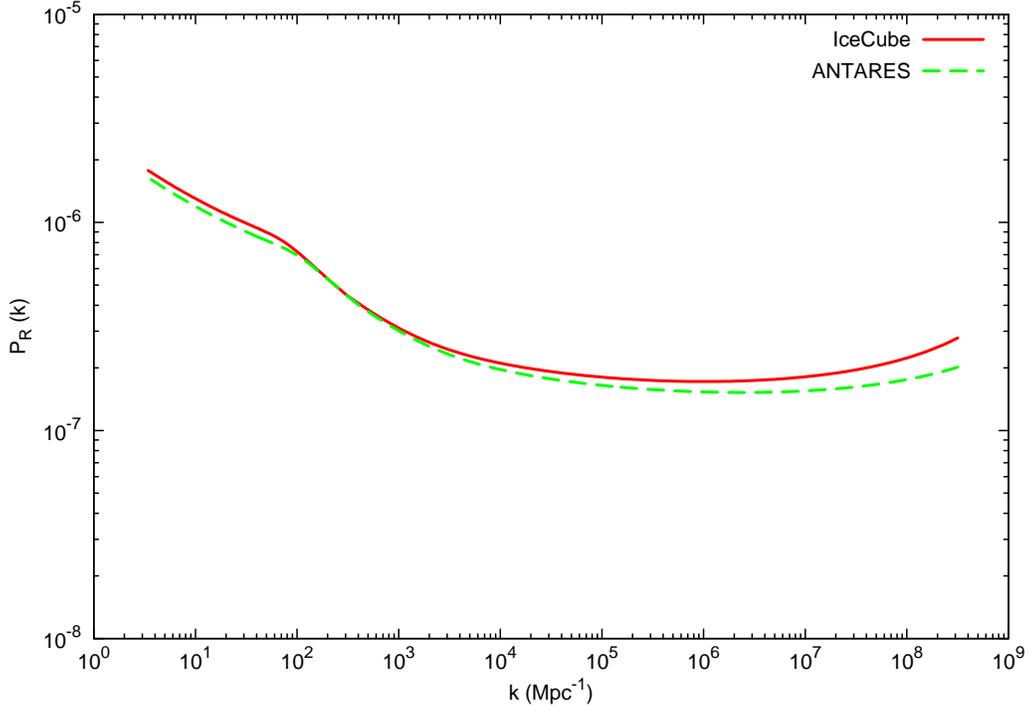,width=0.8\textwidth}
\caption{Upper limits ($95\%$ C.L.) on the amplitude of the primordial curvature perturbations, 
$\mathcal{P}_{\mathcal{R}}(k)$, for 
scales $3 \lesssim k \lesssim 4\times 10^{8} \ \rm Mpc^{-1}$, 
for IceCube (red solid line) 
and ANTARES (green dashed line). 
The parameters of the dark matter particle 
are the same as Fig.~\ref{fraction_plot}.}
\label{cons}
\end{figure}

The constraints on the mass fraction of UCMHs can be used to get the limits 
on the amplitude of the primordial curvature perturbations, $\mathcal{P}_{\mathcal{R}}(k)$. 
Here we briefly review the main processes of calculations and one can 
refer to Refs.~\cite{Bringmann_1,yyp_neutrino} for more detailed discussions. 
As mentioned above, UCMHs can be formed if the early density perturbations  
are in the range of $0.001 \lesssim \delta \rho/\rho \lesssim 0.3$. 
If the initial perturbations are Gaussian, the fraction of UCMHs 
is related to the primordial density perturbations as 

\beqa
\Omega_{\rm UCMHs}=\frac{2\Omega_{\rm DM}}{\sqrt{2\pi}\sigma_{H}(R)}
\frac{M_{\rm UCMHs}(z=0)}{M_{\rm UCMHs}(z=z_{eq})}\int^{\delta_{\rm max}}_{\delta_{\rm min}}
\mathrm{exp}\left(-\frac{\delta^{2}_{H}(R)}{2\delta^{2}_{H}(R)}\right)
d\sigma_{H}(R),
\eeqa
where $\delta_{\rm max}$ and $\delta_{\rm max}$ are the maximal 
and minimal values of density perturbations required for the formation of 
UCMHs and both of them depend on the redshift~\cite{Bringmann_1}. 
$\sigma_{H}(R)$ is related to the curvature perturbations as

\beqa
\sigma^{2}_{H}(R)=\frac{1}{9}\int^{\infty}_{0}x^{3}W^{2}(x)\mathcal{P}_{\mathcal{R}}(x/R)
T^{2}(x/\sqrt{3})dx,
\eeqa
where $W(x)=3x^{-3}(sinx-xcosx)$ is the Fourier transform of the top-hat windows function 
with $x\equiv kR$. $T$ is the transfer function describing the evolution of 
perturbations. Using the above equations, one can translate the limits on the mass fraction 
of UCMHs into the constraints on the amplitude of primordial curvature 
perturbations. The results are shown in Fig.~\ref{cons}. 
From this plot, one can find that the limits on 
the amplitude of primordial curvature perturbations are 
$\mathcal{P}_{\mathcal{R}}(k) \lesssim 2 \times 10^{-6}$ for the scales 
$3 \lesssim k \lesssim 4 \times 10^{8}\ \rm Mpc^{-1}$. The results are comparable 
to that of previous works. The constraints on $f_{\rm UCMHs}$ 
for ANTARES are better that of IceCube, 
therefore, as shown in Fig.~\ref{cons}, the limits on $\mathcal{P}_{\mathcal{R}}(k)$ 
are also better for the ANTARES.

\begin{figure}
\epsfig{file=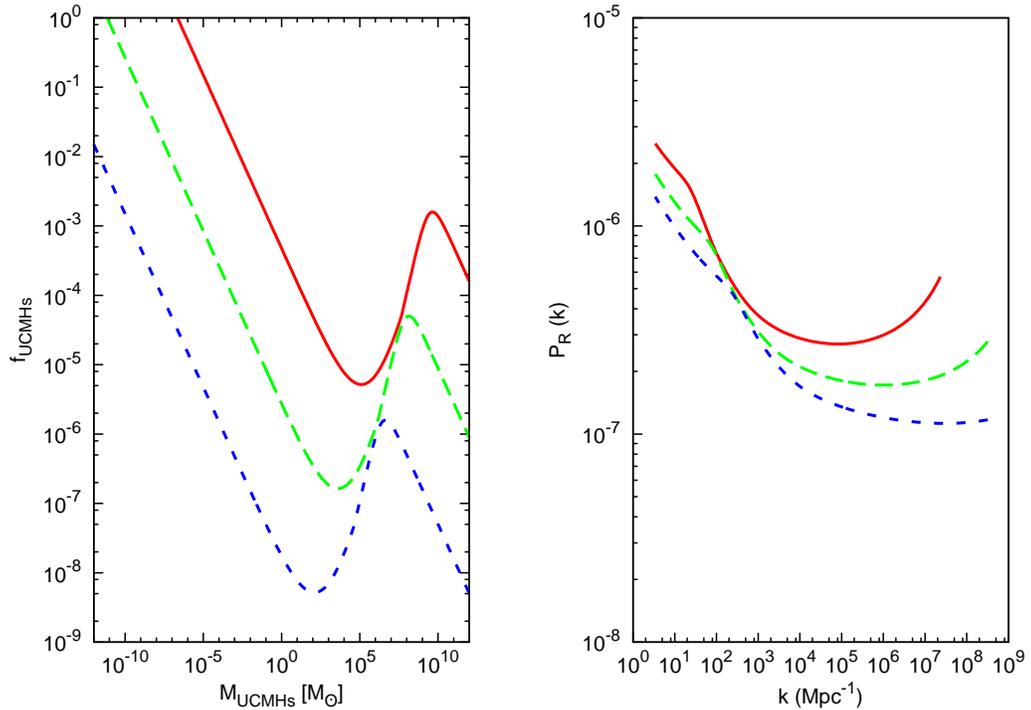,width=0.8\textwidth}
\caption{Constraints on the mass fraction of UCMHs (left) and 
the amplitude of primordial curvature perturbations for the different center density 
profile (right), $r_{\rm min}/R_{\rm UCMH} = 10^{-5}$ (green long-dashed line), 
$10^{-6}$ (blue short-dashed line) and $10^{-7}$ (blue solid line). 
The parameters of the dark matter particle 
are the same as Fig.~\ref{fraction_plot}. 
Here we show the results for IceCube.}
\label{compare}
\end{figure}

\begin{figure}
\epsfig{file=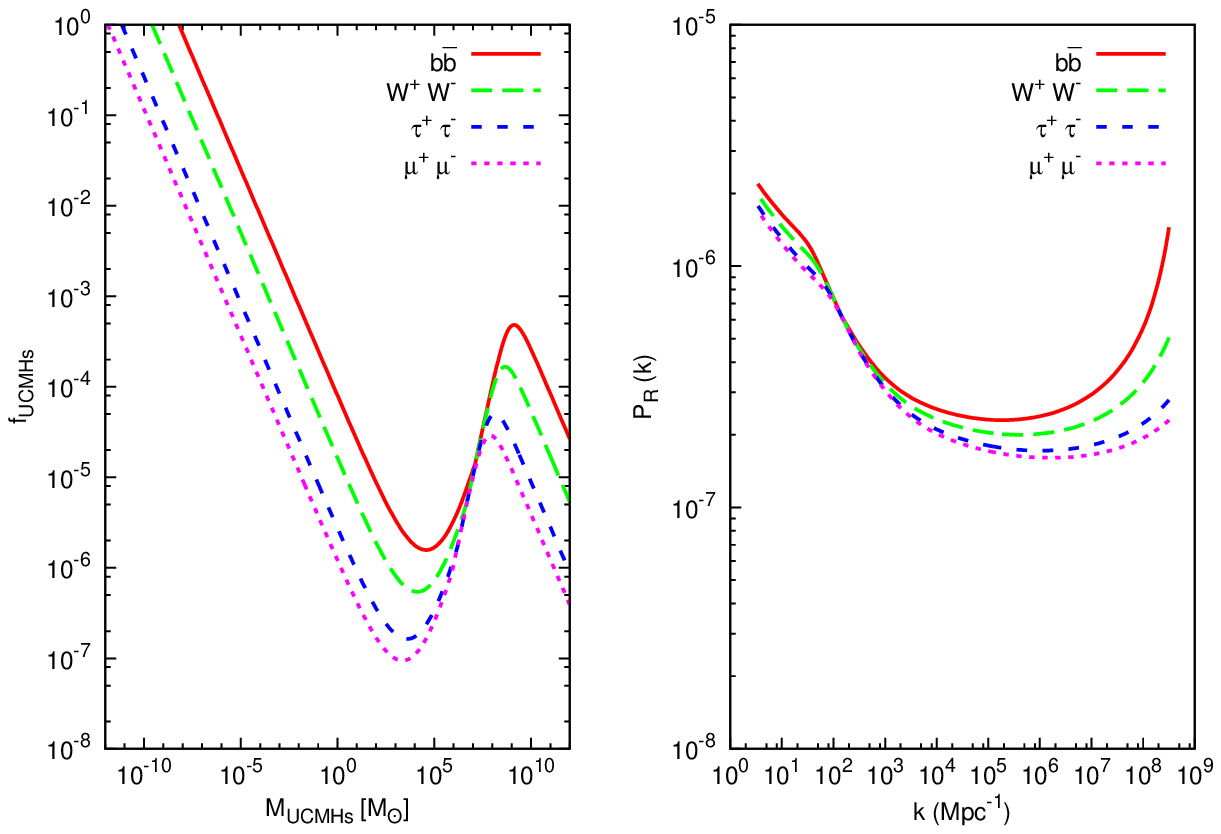,width=0.8\textwidth}
\caption{Constraints on the mass fraction of UCMHs (left) and 
the amplitude of primordial curvature perturbations for different annihilation channels (right), 
$b \bar b$ (red solid line), $W^{+}W^{-}$ (green long-dashed line), 
$\tau^{+}\tau^{-}$ (blue short-dashed line), $\mu^{+}\mu^{-}$ (purple dotted line). 
The parameters of the dark matter particle 
are the same as Fig.~\ref{fraction_plot}. 
For simplicity, we show the results for IceCube.}
\label{compare_for_chan}
\end{figure}

There are several factors 
that can influence the final constraints. One is the inner density profile of the UCMH. 
Since the annihilation rate of the dark matter 
particles is proportional to the square number density, the inner 
density profile of the UCMH is very important for the production of neutrino 
flux caused by dark matter annihilation. The detailed discussions about this issue 
are given in Ref.~\cite{ucmhs_inner}. In this work, for the center density profile, we have 
used Eq.~(\ref{rho_cen}) for our calculations and it is the result of 
dark matter annihilation. The main flaw of using Eq.~(\ref{rho_cen}) is that it neglects 
the infalling of dark matter particles 
after the annihilation~\cite{ucmhs_inner}. Figure~\ref{compare} shows the constraints on the fraction of UCMHs 
and amplitude of primordial curvature perturbations for different $r_{\rm min}$ 
of UCMH. In this plot, we have simply set $r_{\rm min}/R_{\rm UCMH} = 10^{-5}, 10^{-6}$ 
and $10^{-7}$. For the constraints on the mass fraction of UCMHs, there are 
about 2 order differences for some mass range of UCMHs. There are also 
clear differences for the constraints 
on the amplitude of primordial curvature perturbations. 
Another very important 
factor that can affect the final constraints is the misidentification of events 
and the detector efficiency for tau leptons. Detailed discussions 
are given in Ref.~\cite{tau-downward}. As shown in Ref.~\cite{tau-downward}, 
for the choice of reasonable parameters, 
compared with the case of no misidentification, 
the final constraints are about 1 order weaker.\footnote{For detailed discussions 
one can refer to Ref.~\cite{tau-downward}, e.g. the Fig.~10.} 
Besides these two factors, different dark matter 
annihilation channels also have significant impacts on the final results. 
In this paper, we have investigated four annihilation channels, 
$b \bar b, W^{+}W^{-}, \tau^{+}\tau^{-}$ and $\mu^{+}\mu^{-}$. The limits on 
$f_{\rm UCMHs}$ and $\mathcal{P}_{\mathcal{R}}(k)$ for different channels 
are shown in Fig.~\ref{compare_for_chan}. From this plot, it can be seen that 
the better constraints are for the lepton channels, $\mu^{+}\mu^{-}$ and 
$\tau^{+}\tau^{-}$. 
The constraints on $f_{\rm UCMHs}$ are about 2 orders better for 
the $\mu^{+}\mu^{-}$ channel than that of the $b\bar b$ channel for some 
mass range of UCMHs. 
For the $\tau^{+}\tau^{-}$ channel, the limits on $f_{\rm UCMHs}$ are about a factor of 2 weaker
than that of the $\mu^{+}\mu^{-}$ channel. Similar results can also 
be found for the limits on $\mathcal{P}_{\mathcal{R}}(k)$.

\section{conclusions}

In this work, we have investigated the potential downward-going tau neutrino flux 
from UCMHs due to dark matter annihilation. Compared with muon neutrino flux the background 
of tau neutrino flux is smaller while the tau neutrino flux from UCMHs is the same order 
with the muon neutrino flux. With no detection of neutrino flux from UCMHs, we got 
the constraints on the mass fraction of UCMHs. The strongest limits are 
$f_{\rm UCMHs} \lesssim 10^{-7}$ for the mass $M_{\rm UCMH} \sim 10^{4} M_\odot$. 
These results are comparable with previous works. 
Using the limits on the fraction of UCMHs we then got the constraints on the 
amplitude of primordial curvature perturbations on the scales of 
$3 \lesssim k \lesssim 4\times 10^{8}\ \rm Mpc^{-1}$. The strongest limits are 
$\mathcal{P}_{\mathcal{R}}(k) \lesssim 1.5 \times 10^{-7}$ at scale $k \sim 10^{6}\rm\  Mpc^{-1}$. 
Compared with previous works, e.g. Ref.~\cite{yyp_neutrino}, the strongest constraints on 
$\mathcal{P}_{\mathcal{R}}(k)$ are comparable.\footnote{In Ref.~\cite{yyp_neutrino}, 
for the final constraints on $\mathcal{P}_{\mathcal{R}}(k)$, the authors 
have set $\delta_{min}=0.001$ for all scales to get the conservative constraints. 
In fact, the values of $\delta_{\min}$ depend on the redshift~\cite{Bringmann_1}. 
In this work, following the methods in Ref.~\cite{Bringmann_1}, we used the redshift dependent values 
of $\delta_{min}(z)$ for our calculations.} In Ref.~\cite{yyp_neutrino}, the authors used the 
muon neutrino flux to get the constraints on $\mathcal{P}_{\mathcal{R}}(k)$ 
on scales $1 \lesssim k \lesssim 10^{7}\ \rm Mpc^{-1}$. In this work, since the lower background 
of tau neutrino flux, the scales can be extended 
to $k \sim 10^{8}\ \rm Mpc^{-1}$. 

The dark matter annihilation rate is proportional to the square number density 
of dark matter particles, therefore, the center density profile of UCMHs 
is very important for the constraints on $f_{\rm UCMHs}$ and $\mathcal{P}_{\mathcal{R}}(k)$. 
The research on the center density profile of the UCMH is beyond the scope of this work and 
detailed discussions on this issue can be found in Ref.~\cite{ucmhs_inner}. In this paper, 
in order to investigate the influences of different center density profile on the final constraints 
we simply considered three forms of center density profile of the UCMH. 
Specifically, we have set $r_{\rm min}/R_{\rm UCMH} = 10^{-5}, 10^{-6}$ 
and $10^{-7}$ for the purpose. For these settings, there are about 2 orders differences for the constraints 
on $f_{\rm UCMHs}$ and also obvious differences for the constraints on $\mathcal{P}_{\mathcal{R}}(k)$. 
Another factor that can effect the finals results is the misidentification of events. 
Detailed discussions can be found in Ref.~\cite{tau-downward}. According to their calculations, 
for the choice of reasonable parameters, there is about 1 order 
difference for the final constraints. 
Besides the above factors, different dark matter annihilation channels 
have also significant impacts on the final constraints. In order to investigate 
these impacts, we have set four 
channels for our calculations, $b \bar b, W^{+}W^{-}, \tau^{+}\tau^{-}$ and $\mu^{+}\mu^{-}$. 
We have found that for the limits on $f_{\rm UCMHs}$ the best results are from the $\mu^{+}\mu^{-}$ channel. 
There are no big differences between $\mu^{+}\mu^{-}$ and 
$\tau^{+}\tau^{-}$ channels. 
Similar results can also be found for the limits on $\mathcal{P}_{\mathcal{R}}(k)$.

\acknowledgments

We thank Q. Yuan for very useful discussions and the anonymous referee for a very useful report. 
Yupeng Yang is supported by the National Science Foundation
of China (Grants No.U1404114 and No.11505005).


\begin{thebibliography}{53}%
\makeatletter
\providecommand \@ifxundefined [1]{%
 \@ifx{#1\undefined}
}%
\providecommand \@ifnum [1]{%
 \ifnum #1\expandafter \@firstoftwo
 \else \expandafter \@secondoftwo
 \fi
}%
\providecommand \@ifx [1]{%
 \ifx #1\expandafter \@firstoftwo
 \else \expandafter \@secondoftwo
 \fi
}%
\providecommand \natexlab [1]{#1}%
\providecommand \enquote  [1]{``#1''}%
\providecommand \bibnamefont  [1]{#1}%
\providecommand \bibfnamefont [1]{#1}%
\providecommand \citenamefont [1]{#1}%
\providecommand \href@noop [0]{\@secondoftwo}%
\providecommand \href [0]{\begingroup \@sanitize@url \@href}%
\providecommand \@href[1]{\@@startlink{#1}\@@href}%
\providecommand \@@href[1]{\endgroup#1\@@endlink}%
\providecommand \@sanitize@url [0]{\catcode `\\12\catcode `\$12\catcode
  `\&12\catcode `\#12\catcode `\^12\catcode `\_12\catcode `\%12\relax}%
\providecommand \@@startlink[1]{}%
\providecommand \@@endlink[0]{}%
\providecommand \url  [0]{\begingroup\@sanitize@url \@url }%
\providecommand \@url [1]{\endgroup\@href {#1}{\urlprefix }}%
\providecommand \urlprefix  [0]{URL }%
\providecommand \Eprint [0]{\href }%
\providecommand \doibase [0]{http://dx.doi.org/}%
\providecommand \selectlanguage [0]{\@gobble}%
\providecommand \bibinfo  [0]{\@secondoftwo}%
\providecommand \bibfield  [0]{\@secondoftwo}%
\providecommand \translation [1]{[#1]}%
\providecommand \BibitemOpen [0]{}%
\providecommand \bibitemStop [0]{}%
\providecommand \bibitemNoStop [0]{.\EOS\space}%
\providecommand \EOS [0]{\spacefactor3000\relax}%
\providecommand \BibitemShut  [1]{\csname bibitem#1\endcsname}%
\let\auto@bib@innerbib\@empty
\bibitem [{\citenamefont {Jungman}\ \emph {et~al.}(1996)\citenamefont
  {Jungman}, \citenamefont {Kamionkowski},\ and\ \citenamefont
  {Griest}}]{darkmatter_1}%
  \BibitemOpen
  \bibfield  {author} {\bibinfo {author} {\bibfnamefont {G.}~\bibnamefont
  {Jungman}}, \bibinfo {author} {\bibfnamefont {M.}~\bibnamefont
  {Kamionkowski}}, \ and\ \bibinfo {author} {\bibfnamefont {K.}~\bibnamefont
  {Griest}},\ }\href {\doibase 10.1016/0370-1573(95)00058-5} {\bibfield
  {journal} {\bibinfo  {journal} {Phys. Rept.}\ }\textbf {\bibinfo {volume}
  {267}},\ \bibinfo {pages} {195} (\bibinfo {year} {1996})},\ \Eprint
  {http://arxiv.org/abs/hep-ph/9506380} {arXiv:hep-ph/9506380 [hep-ph]}
  \BibitemShut {NoStop}%
\bibitem [{\citenamefont {Bertone}\ \emph {et~al.}(2005)\citenamefont
  {Bertone}, \citenamefont {Hooper},\ and\ \citenamefont
  {Silk}}]{darkmatter_2}%
  \BibitemOpen
  \bibfield  {author} {\bibinfo {author} {\bibfnamefont {G.}~\bibnamefont
  {Bertone}}, \bibinfo {author} {\bibfnamefont {D.}~\bibnamefont {Hooper}}, \
  and\ \bibinfo {author} {\bibfnamefont {J.}~\bibnamefont {Silk}},\ }\href
  {\doibase 10.1016/j.physrep.2004.08.031} {\bibfield  {journal} {\bibinfo
  {journal} {Phys. Rept.}\ }\textbf {\bibinfo {volume} {405}},\ \bibinfo
  {pages} {279} (\bibinfo {year} {2005})},\ \Eprint
  {http://arxiv.org/abs/hep-ph/0404175} {arXiv:hep-ph/0404175 [hep-ph]}
  \BibitemShut {NoStop}%
\bibitem [{\citenamefont {Gaskins}(2016)}]{dm_cons_1}%
  \BibitemOpen
  \bibfield  {author} {\bibinfo {author} {\bibfnamefont {J.~M.}\ \bibnamefont
  {Gaskins}},\ }\href {\doibase 10.1080/00107514.2016.1175160} {\bibfield
  {journal} {\bibinfo  {journal} {Contemp. Phys.}\ }\textbf {\bibinfo {volume}
  {57}},\ \bibinfo {pages} {496} (\bibinfo {year} {2016})},\ \Eprint
  {http://arxiv.org/abs/1604.00014} {arXiv:1604.00014 [astro-ph.HE]}
  \BibitemShut {NoStop}%
\bibitem [{\citenamefont {Ackermann}\ \emph {et~al.}(2015)\citenamefont
  {Ackermann} \emph {et~al.}}]{dm_cons_2}%
  \BibitemOpen
  \bibfield  {author} {\bibinfo {author} {\bibfnamefont {M.}~\bibnamefont
  {Ackermann}} \emph {et~al.} (\bibinfo {collaboration} {Fermi-LAT}),\ }\href
  {\doibase 10.1088/1475-7516/2015/09/008} {\bibfield  {journal} {\bibinfo
  {journal} {JCAP}\ }\textbf {\bibinfo {volume} {1509}},\ \bibinfo {pages}
  {008} (\bibinfo {year} {2015})},\ \Eprint {http://arxiv.org/abs/1501.05464}
  {arXiv:1501.05464 [astro-ph.CO]} \BibitemShut {NoStop}%
\bibitem [{\citenamefont {Ackermann}\ \emph {et~al.}(2011)\citenamefont
  {Ackermann} \emph {et~al.}}]{dm_cons_3}%
  \BibitemOpen
  \bibfield  {author} {\bibinfo {author} {\bibfnamefont {M.}~\bibnamefont
  {Ackermann}} \emph {et~al.} (\bibinfo {collaboration} {Fermi-LAT}),\ }\href
  {\doibase 10.1103/PhysRevLett.107.241302} {\bibfield  {journal} {\bibinfo
  {journal} {Phys. Rev. Lett.}\ }\textbf {\bibinfo {volume} {107}},\ \bibinfo
  {pages} {241302} (\bibinfo {year} {2011})},\ \Eprint
  {http://arxiv.org/abs/1108.3546} {arXiv:1108.3546 [astro-ph.HE]} \BibitemShut
  {NoStop}%
\bibitem [{\citenamefont {Liu}\ \emph {et~al.}(2017)\citenamefont {Liu},
  \citenamefont {Bi}, \citenamefont {Lin},\ and\ \citenamefont {Yin}}]{liuwei}%
  \BibitemOpen
  \bibfield  {author} {\bibinfo {author} {\bibfnamefont {W.}~\bibnamefont
  {Liu}}, \bibinfo {author} {\bibfnamefont {X.-J.}\ \bibnamefont {Bi}},
  \bibinfo {author} {\bibfnamefont {S.-J.}\ \bibnamefont {Lin}}, \ and\
  \bibinfo {author} {\bibfnamefont {P.-F.}\ \bibnamefont {Yin}},\ }\href
  {\doibase 10.1088/1674-1137/41/4/045104} {\bibfield  {journal} {\bibinfo
  {journal} {Chin. Phys.}\ }\textbf {\bibinfo {volume} {C41}},\ \bibinfo
  {pages} {045104} (\bibinfo {year} {2017})},\ \Eprint
  {http://arxiv.org/abs/1602.01012} {arXiv:1602.01012 [astro-ph.CO]}
  \BibitemShut {NoStop}%
\bibitem [{\citenamefont {Sandick}\ \emph {et~al.}(2010)\citenamefont
  {Sandick}, \citenamefont {Spolyar}, \citenamefont {Buckley}, \citenamefont
  {Freese},\ and\ \citenamefont {Hooper}}]{dm_cons_neutrino_1}%
  \BibitemOpen
  \bibfield  {author} {\bibinfo {author} {\bibfnamefont {P.}~\bibnamefont
  {Sandick}}, \bibinfo {author} {\bibfnamefont {D.}~\bibnamefont {Spolyar}},
  \bibinfo {author} {\bibfnamefont {M.}~\bibnamefont {Buckley}}, \bibinfo
  {author} {\bibfnamefont {K.}~\bibnamefont {Freese}}, \ and\ \bibinfo {author}
  {\bibfnamefont {D.}~\bibnamefont {Hooper}},\ }\href {\doibase
  10.1103/PhysRevD.81.083506} {\bibfield  {journal} {\bibinfo  {journal} {Phys.
  Rev. D}\ }\textbf {\bibinfo {volume} {81}},\ \bibinfo {pages} {083506}
  (\bibinfo {year} {2010})}\BibitemShut {NoStop}%
\bibitem [{\citenamefont {{The IceCube
  Collaboration}}(2011)}]{dm_cons_neutrino_2}%
  \BibitemOpen
  \bibfield  {author} {\bibinfo {author} {\bibnamefont {{The IceCube
  Collaboration}}},\ }\href@noop {} {\bibfield  {journal} {\bibinfo  {journal}
  {ArXiv e-prints}\ } (\bibinfo {year} {2011})},\ \Eprint
  {http://arxiv.org/abs/1111.2738} {arXiv:1111.2738 [astro-ph.HE]} \BibitemShut
  {NoStop}%
\bibitem [{\citenamefont {Adrian-Martinez}\ \emph {et~al.}(2015)\citenamefont
  {Adrian-Martinez} \emph {et~al.}}]{Adrian-Martinez:2015wey}%
  \BibitemOpen
  \bibfield  {author} {\bibinfo {author} {\bibfnamefont {S.}~\bibnamefont
  {Adrian-Martinez}} \emph {et~al.} (\bibinfo {collaboration} {ANTARES}),\
  }\href {\doibase 10.1088/1475-7516/2015/10/068} {\bibfield  {journal}
  {\bibinfo  {journal} {JCAP}\ }\textbf {\bibinfo {volume} {1510}},\ \bibinfo
  {pages} {068} (\bibinfo {year} {2015})},\ \Eprint
  {http://arxiv.org/abs/1505.04866} {arXiv:1505.04866 [astro-ph.HE]}
  \BibitemShut {NoStop}%
\bibitem [{\citenamefont {Bottino}\ \emph {et~al.}(2004)\citenamefont
  {Bottino}, \citenamefont {Donato}, \citenamefont {Fornengo},\ and\
  \citenamefont {Scopel}}]{Bottino:2004qi}%
  \BibitemOpen
  \bibfield  {author} {\bibinfo {author} {\bibfnamefont {A.}~\bibnamefont
  {Bottino}}, \bibinfo {author} {\bibfnamefont {F.}~\bibnamefont {Donato}},
  \bibinfo {author} {\bibfnamefont {N.}~\bibnamefont {Fornengo}}, \ and\
  \bibinfo {author} {\bibfnamefont {S.}~\bibnamefont {Scopel}},\ }\href
  {\doibase 10.1103/PhysRevD.70.015005} {\bibfield  {journal} {\bibinfo
  {journal} {Phys. Rev.}\ }\textbf {\bibinfo {volume} {D70}},\ \bibinfo {pages}
  {015005} (\bibinfo {year} {2004})},\ \Eprint
  {http://arxiv.org/abs/hep-ph/0401186} {arXiv:hep-ph/0401186 [hep-ph]}
  \BibitemShut {NoStop}%
\bibitem [{\citenamefont {Bottino}\ \emph {et~al.}(1995)\citenamefont
  {Bottino}, \citenamefont {Fornengo}, \citenamefont {Mignola},\ and\
  \citenamefont {Moscoso}}]{Bottino:1994xp}%
  \BibitemOpen
  \bibfield  {author} {\bibinfo {author} {\bibfnamefont {A.}~\bibnamefont
  {Bottino}}, \bibinfo {author} {\bibfnamefont {N.}~\bibnamefont {Fornengo}},
  \bibinfo {author} {\bibfnamefont {G.}~\bibnamefont {Mignola}}, \ and\
  \bibinfo {author} {\bibfnamefont {L.}~\bibnamefont {Moscoso}},\ }\href
  {\doibase 10.1016/0927-6505(94)00028-2} {\bibfield  {journal} {\bibinfo
  {journal} {Astropart. Phys.}\ }\textbf {\bibinfo {volume} {3}},\ \bibinfo
  {pages} {65} (\bibinfo {year} {1995})},\ \Eprint
  {http://arxiv.org/abs/hep-ph/9408391} {arXiv:hep-ph/9408391 [hep-ph]}
  \BibitemShut {NoStop}%
\bibitem [{\citenamefont {Blennow}\ \emph {et~al.}(2013)\citenamefont
  {Blennow}, \citenamefont {Carrigan},\ and\ \citenamefont
  {Fernandez~Martinez}}]{Blennow:2013pya}%
  \BibitemOpen
  \bibfield  {author} {\bibinfo {author} {\bibfnamefont {M.}~\bibnamefont
  {Blennow}}, \bibinfo {author} {\bibfnamefont {M.}~\bibnamefont {Carrigan}}, \
  and\ \bibinfo {author} {\bibfnamefont {E.}~\bibnamefont
  {Fernandez~Martinez}},\ }\href {\doibase 10.1088/1475-7516/2013/06/038}
  {\bibfield  {journal} {\bibinfo  {journal} {JCAP}\ }\textbf {\bibinfo
  {volume} {1306}},\ \bibinfo {pages} {038} (\bibinfo {year} {2013})},\ \Eprint
  {http://arxiv.org/abs/1303.4530} {arXiv:1303.4530 [hep-ph]} \BibitemShut
  {NoStop}%
\bibitem [{\citenamefont {Fornengo}\ and\ \citenamefont
  {Niro}(2011)}]{tau-downward}%
  \BibitemOpen
  \bibfield  {author} {\bibinfo {author} {\bibfnamefont {N.}~\bibnamefont
  {Fornengo}}\ and\ \bibinfo {author} {\bibfnamefont {V.}~\bibnamefont
  {Niro}},\ }\href {\doibase 10.1007/JHEP11(2011)133} {\bibfield  {journal}
  {\bibinfo  {journal} {JHEP}\ }\textbf {\bibinfo {volume} {11}},\ \bibinfo
  {pages} {133} (\bibinfo {year} {2011})},\ \Eprint
  {http://arxiv.org/abs/1108.2630} {arXiv:1108.2630 [hep-ph]} \BibitemShut
  {NoStop}%
\bibitem [{\citenamefont {Covi}\ \emph {et~al.}(2009)\citenamefont {Covi},
  \citenamefont {Grefe}, \citenamefont {Ibarra},\ and\ \citenamefont
  {Tran}}]{dm_decay_neutrino}%
  \BibitemOpen
  \bibfield  {author} {\bibinfo {author} {\bibfnamefont {L.}~\bibnamefont
  {Covi}}, \bibinfo {author} {\bibfnamefont {M.}~\bibnamefont {Grefe}},
  \bibinfo {author} {\bibfnamefont {A.}~\bibnamefont {Ibarra}}, \ and\ \bibinfo
  {author} {\bibfnamefont {D.}~\bibnamefont {Tran}},\ }\href {\doibase
  10.1088/1475-7516/2009/01/029} {\bibfield  {journal} {\bibinfo  {journal}
  {JCAP}\ }\textbf {\bibinfo {volume} {0901}},\ \bibinfo {pages} {029}
  (\bibinfo {year} {2009})},\ \Eprint {http://arxiv.org/abs/0809.5030}
  {arXiv:0809.5030 [hep-ph]} \BibitemShut {NoStop}%
\bibitem [{\citenamefont {Kumar}\ \emph {et~al.}(2011)\citenamefont {Kumar},
  \citenamefont {Learned}, \citenamefont {Sakai},\ and\ \citenamefont
  {Smith}}]{Kumar:2011hi}%
  \BibitemOpen
  \bibfield  {author} {\bibinfo {author} {\bibfnamefont {J.}~\bibnamefont
  {Kumar}}, \bibinfo {author} {\bibfnamefont {J.~G.}\ \bibnamefont {Learned}},
  \bibinfo {author} {\bibfnamefont {M.}~\bibnamefont {Sakai}}, \ and\ \bibinfo
  {author} {\bibfnamefont {S.}~\bibnamefont {Smith}},\ }\href {\doibase
  10.1103/PhysRevD.84.036007} {\bibfield  {journal} {\bibinfo  {journal} {Phys.
  Rev.}\ }\textbf {\bibinfo {volume} {D84}},\ \bibinfo {pages} {036007}
  (\bibinfo {year} {2011})},\ \Eprint {http://arxiv.org/abs/1103.3270}
  {arXiv:1103.3270 [hep-ph]} \BibitemShut {NoStop}%
\bibitem [{\citenamefont {Guo}(2016)}]{Guo:2015hsy}%
  \BibitemOpen
  \bibfield  {author} {\bibinfo {author} {\bibfnamefont {W.-L.}\ \bibnamefont
  {Guo}},\ }\href {\doibase 10.1088/1475-7516/2016/01/039} {\bibfield
  {journal} {\bibinfo  {journal} {JCAP}\ }\textbf {\bibinfo {volume} {1601}},\
  \bibinfo {pages} {039} (\bibinfo {year} {2016})},\ \Eprint
  {http://arxiv.org/abs/1511.04888} {arXiv:1511.04888 [hep-ph]} \BibitemShut
  {NoStop}%
\bibitem [{\citenamefont {Xu}(2016)}]{Xu:2017yxo}%
  \BibitemOpen
  \bibfield  {author} {\bibinfo {author} {\bibfnamefont {D.}~\bibnamefont {Xu}}
  (\bibinfo {collaboration} {IceCube}),\ }\bibfield  {booktitle} {\emph
  {\bibinfo {booktitle} {{Proceedings, 38th International Conference on High
  Energy Physics (ICHEP 2016): Chicago, IL, USA, August 3-10, 2016}}},\
  }\href@noop {} {\bibfield  {journal} {\bibinfo  {journal} {PoS}\ }\textbf
  {\bibinfo {volume} {ICHEP2016}},\ \bibinfo {pages} {452} (\bibinfo {year}
  {2016})},\ \Eprint {http://arxiv.org/abs/1702.05238} {arXiv:1702.05238
  [astro-ph.HE]} \BibitemShut {NoStop}%
\bibitem [{\citenamefont {Conrad}\ \emph {et~al.}(2010)\citenamefont {Conrad},
  \citenamefont {de~Gouvea}, \citenamefont {Shalgar},\ and\ \citenamefont
  {Spitz}}]{Conrad:2010mh}%
  \BibitemOpen
  \bibfield  {author} {\bibinfo {author} {\bibfnamefont {J.}~\bibnamefont
  {Conrad}}, \bibinfo {author} {\bibfnamefont {A.}~\bibnamefont {de~Gouvea}},
  \bibinfo {author} {\bibfnamefont {S.}~\bibnamefont {Shalgar}}, \ and\
  \bibinfo {author} {\bibfnamefont {J.}~\bibnamefont {Spitz}},\ }\href
  {\doibase 10.1103/PhysRevD.82.093012} {\bibfield  {journal} {\bibinfo
  {journal} {Phys. Rev.}\ }\textbf {\bibinfo {volume} {D82}},\ \bibinfo {pages}
  {093012} (\bibinfo {year} {2010})},\ \Eprint {http://arxiv.org/abs/1008.2984}
  {arXiv:1008.2984 [hep-ph]} \BibitemShut {NoStop}%
\bibitem [{\citenamefont {Cuoco}\ \emph {et~al.}(2007)\citenamefont {Cuoco},
  \citenamefont {Mangano}, \citenamefont {Miele}, \citenamefont {Pastor},
  \citenamefont {Perrone}, \citenamefont {Pisanti},\ and\ \citenamefont
  {Serpico}}]{Cuoco:2006qd}%
  \BibitemOpen
  \bibfield  {author} {\bibinfo {author} {\bibfnamefont {A.}~\bibnamefont
  {Cuoco}}, \bibinfo {author} {\bibfnamefont {G.}~\bibnamefont {Mangano}},
  \bibinfo {author} {\bibfnamefont {G.}~\bibnamefont {Miele}}, \bibinfo
  {author} {\bibfnamefont {S.}~\bibnamefont {Pastor}}, \bibinfo {author}
  {\bibfnamefont {L.}~\bibnamefont {Perrone}}, \bibinfo {author} {\bibfnamefont
  {O.}~\bibnamefont {Pisanti}}, \ and\ \bibinfo {author} {\bibfnamefont
  {P.~D.}\ \bibnamefont {Serpico}},\ }\href {\doibase
  10.1088/1475-7516/2007/02/007} {\bibfield  {journal} {\bibinfo  {journal}
  {JCAP}\ }\textbf {\bibinfo {volume} {0702}},\ \bibinfo {pages} {007}
  (\bibinfo {year} {2007})},\ \Eprint {http://arxiv.org/abs/astro-ph/0609241}
  {arXiv:astro-ph/0609241 [astro-ph]} \BibitemShut {NoStop}%
\bibitem [{\citenamefont {Hlozek}\ \emph {et~al.}(2012)\citenamefont {Hlozek},
  \citenamefont {Dunkley}, \citenamefont {Addison}, \citenamefont {Appel},
  \citenamefont {Bond}, \citenamefont {Carvalho}, \citenamefont {Das},
  \citenamefont {Devlin}, \citenamefont {Dünner}, \citenamefont
  {Essinger-Hileman}, \citenamefont {Fowler}, \citenamefont {Gallardo},
  \citenamefont {Hajian}, \citenamefont {Halpern}, \citenamefont {Hasselfield},
  \citenamefont {Hilton}, \citenamefont {Hincks}, \citenamefont {Hughes},
  \citenamefont {Irwin}, \citenamefont {Klein}, \citenamefont {Kosowsky},
  \citenamefont {Marriage}, \citenamefont {Marsden}, \citenamefont {Menanteau},
  \citenamefont {Moodley}, \citenamefont {Niemack}, \citenamefont {Nolta},
  \citenamefont {Page}, \citenamefont {Parker}, \citenamefont {Partridge},
  \citenamefont {Rojas}, \citenamefont {Sehgal}, \citenamefont {Sherwin},
  \citenamefont {Sievers}, \citenamefont {Spergel}, \citenamefont {Staggs},
  \citenamefont {Swetz}, \citenamefont {Switzer}, \citenamefont {Thornton},\
  and\ \citenamefont {Wollack}}]{cmb_2}%
  \BibitemOpen
  \bibfield  {author} {\bibinfo {author} {\bibfnamefont {R.}~\bibnamefont
  {Hlozek}}, \bibinfo {author} {\bibfnamefont {J.}~\bibnamefont {Dunkley}},
  \bibinfo {author} {\bibfnamefont {G.}~\bibnamefont {Addison}}, \bibinfo
  {author} {\bibfnamefont {J.~W.}\ \bibnamefont {Appel}}, \bibinfo {author}
  {\bibfnamefont {J.~R.}\ \bibnamefont {Bond}}, \bibinfo {author}
  {\bibfnamefont {C.~S.}\ \bibnamefont {Carvalho}}, \bibinfo {author}
  {\bibfnamefont {S.}~\bibnamefont {Das}}, \bibinfo {author} {\bibfnamefont
  {M.~J.}\ \bibnamefont {Devlin}}, \bibinfo {author} {\bibfnamefont
  {R.}~\bibnamefont {Dünner}}, \bibinfo {author} {\bibfnamefont
  {T.}~\bibnamefont {Essinger-Hileman}}, \bibinfo {author} {\bibfnamefont
  {J.~W.}\ \bibnamefont {Fowler}}, \bibinfo {author} {\bibfnamefont
  {P.}~\bibnamefont {Gallardo}}, \bibinfo {author} {\bibfnamefont
  {A.}~\bibnamefont {Hajian}}, \bibinfo {author} {\bibfnamefont
  {M.}~\bibnamefont {Halpern}}, \bibinfo {author} {\bibfnamefont
  {M.}~\bibnamefont {Hasselfield}}, \bibinfo {author} {\bibfnamefont
  {M.}~\bibnamefont {Hilton}}, \bibinfo {author} {\bibfnamefont {A.~D.}\
  \bibnamefont {Hincks}}, \bibinfo {author} {\bibfnamefont {J.~P.}\
  \bibnamefont {Hughes}}, \bibinfo {author} {\bibfnamefont {K.~D.}\
  \bibnamefont {Irwin}}, \bibinfo {author} {\bibfnamefont {J.}~\bibnamefont
  {Klein}}, \bibinfo {author} {\bibfnamefont {A.}~\bibnamefont {Kosowsky}},
  \bibinfo {author} {\bibfnamefont {T.~A.}\ \bibnamefont {Marriage}}, \bibinfo
  {author} {\bibfnamefont {D.}~\bibnamefont {Marsden}}, \bibinfo {author}
  {\bibfnamefont {F.}~\bibnamefont {Menanteau}}, \bibinfo {author}
  {\bibfnamefont {K.}~\bibnamefont {Moodley}}, \bibinfo {author} {\bibfnamefont
  {M.~D.}\ \bibnamefont {Niemack}}, \bibinfo {author} {\bibfnamefont {M.~R.}\
  \bibnamefont {Nolta}}, \bibinfo {author} {\bibfnamefont {L.~A.}\ \bibnamefont
  {Page}}, \bibinfo {author} {\bibfnamefont {L.}~\bibnamefont {Parker}},
  \bibinfo {author} {\bibfnamefont {B.}~\bibnamefont {Partridge}}, \bibinfo
  {author} {\bibfnamefont {F.}~\bibnamefont {Rojas}}, \bibinfo {author}
  {\bibfnamefont {N.}~\bibnamefont {Sehgal}}, \bibinfo {author} {\bibfnamefont
  {B.}~\bibnamefont {Sherwin}}, \bibinfo {author} {\bibfnamefont
  {J.}~\bibnamefont {Sievers}}, \bibinfo {author} {\bibfnamefont {D.~N.}\
  \bibnamefont {Spergel}}, \bibinfo {author} {\bibfnamefont {S.~T.}\
  \bibnamefont {Staggs}}, \bibinfo {author} {\bibfnamefont {D.~S.}\
  \bibnamefont {Swetz}}, \bibinfo {author} {\bibfnamefont {E.~R.}\ \bibnamefont
  {Switzer}}, \bibinfo {author} {\bibfnamefont {R.}~\bibnamefont {Thornton}}, \
  and\ \bibinfo {author} {\bibfnamefont {E.}~\bibnamefont {Wollack}},\ }\href
  {http://stacks.iop.org/0004-637X/749/i=1/a=90} {\bibfield  {journal}
  {\bibinfo  {journal} {The Astrophysical Journal}\ }\textbf {\bibinfo {volume}
  {749}},\ \bibinfo {pages} {90} (\bibinfo {year} {2012})}\BibitemShut
  {NoStop}%
\bibitem [{\citenamefont {Green}\ and\ \citenamefont {Liddle}(1997)}]{pbh}%
  \BibitemOpen
  \bibfield  {author} {\bibinfo {author} {\bibfnamefont {A.~M.}\ \bibnamefont
  {Green}}\ and\ \bibinfo {author} {\bibfnamefont {A.~R.}\ \bibnamefont
  {Liddle}},\ }\href {\doibase 10.1103/PhysRevD.56.6166} {\bibfield  {journal}
  {\bibinfo  {journal} {Phys. Rev. D}\ }\textbf {\bibinfo {volume} {56}},\
  \bibinfo {pages} {6166} (\bibinfo {year} {1997})}\BibitemShut {NoStop}%
\bibitem [{\citenamefont {Ricotti}\ and\ \citenamefont
  {Gould}(2009)}]{ucmhs_1}%
  \BibitemOpen
  \bibfield  {author} {\bibinfo {author} {\bibfnamefont {M.}~\bibnamefont
  {Ricotti}}\ and\ \bibinfo {author} {\bibfnamefont {A.}~\bibnamefont
  {Gould}},\ }\href {\doibase 10.1088/0004-637X/707/2/979} {\bibfield
  {journal} {\bibinfo  {journal} {Astrophys. J.}\ }\textbf {\bibinfo {volume}
  {707}},\ \bibinfo {pages} {979} (\bibinfo {year} {2009})},\ \Eprint
  {http://arxiv.org/abs/0908.0735} {arXiv:0908.0735 [astro-ph.CO]} \BibitemShut
  {NoStop}%
\bibitem [{\citenamefont {Navarro}\ \emph {et~al.}(1997)\citenamefont
  {Navarro}, \citenamefont {Frenk},\ and\ \citenamefont {White}}]{NFW}%
  \BibitemOpen
  \bibfield  {author} {\bibinfo {author} {\bibfnamefont {J.~F.}\ \bibnamefont
  {Navarro}}, \bibinfo {author} {\bibfnamefont {C.~S.}\ \bibnamefont {Frenk}},
  \ and\ \bibinfo {author} {\bibfnamefont {S.~D.~M.}\ \bibnamefont {White}},\
  }\href {\doibase 10.1086/304888} {\bibfield  {journal} {\bibinfo  {journal}
  {Astrophys. J.}\ }\textbf {\bibinfo {volume} {490}},\ \bibinfo {pages} {493}
  (\bibinfo {year} {1997})},\ \Eprint {http://arxiv.org/abs/astro-ph/9611107}
  {arXiv:astro-ph/9611107 [astro-ph]} \BibitemShut {NoStop}%
\bibitem [{\citenamefont {Scott}\ and\ \citenamefont
  {Sivertsson}(2009)}]{scott_prl}%
  \BibitemOpen
  \bibfield  {author} {\bibinfo {author} {\bibfnamefont {P.}~\bibnamefont
  {Scott}}\ and\ \bibinfo {author} {\bibfnamefont {S.}~\bibnamefont
  {Sivertsson}},\ }\href {\doibase 10.1103/PhysRevLett.105.119902,
  10.1103/PhysRevLett.103.211301} {\bibfield  {journal} {\bibinfo  {journal}
  {Phys. Rev. Lett.}\ }\textbf {\bibinfo {volume} {103}},\ \bibinfo {pages}
  {211301} (\bibinfo {year} {2009})},\ \bibinfo {note} {[Erratum: Phys. Rev.
  Lett.105,119902(2010)]},\ \Eprint {http://arxiv.org/abs/0908.4082}
  {arXiv:0908.4082 [astro-ph.CO]} \BibitemShut {NoStop}%
\bibitem [{\citenamefont {Yang}\ \emph {et~al.}(2013)\citenamefont {Yang},
  \citenamefont {Yang},\ and\ \citenamefont {Zong}}]{yyp_neutrino}%
  \BibitemOpen
  \bibfield  {author} {\bibinfo {author} {\bibfnamefont {Y.}~\bibnamefont
  {Yang}}, \bibinfo {author} {\bibfnamefont {G.}~\bibnamefont {Yang}}, \ and\
  \bibinfo {author} {\bibfnamefont {H.}~\bibnamefont {Zong}},\ }\href {\doibase
  10.1103/PhysRevD.87.103525} {\bibfield  {journal} {\bibinfo  {journal} {Phys.
  Rev. D}\ }\textbf {\bibinfo {volume} {87}},\ \bibinfo {pages} {103525}
  (\bibinfo {year} {2013})}\BibitemShut {NoStop}%
\bibitem [{\citenamefont {Bringmann}\ \emph {et~al.}(2012)\citenamefont
  {Bringmann}, \citenamefont {Scott},\ and\ \citenamefont
  {Akrami}}]{Bringmann_1}%
  \BibitemOpen
  \bibfield  {author} {\bibinfo {author} {\bibfnamefont {T.}~\bibnamefont
  {Bringmann}}, \bibinfo {author} {\bibfnamefont {P.}~\bibnamefont {Scott}}, \
  and\ \bibinfo {author} {\bibfnamefont {Y.}~\bibnamefont {Akrami}},\ }\href
  {\doibase 10.1103/PhysRevD.85.125027} {\bibfield  {journal} {\bibinfo
  {journal} {Phys. Rev. D}\ }\textbf {\bibinfo {volume} {85}},\ \bibinfo
  {pages} {125027} (\bibinfo {year} {2012})}\BibitemShut {NoStop}%
\bibitem [{\citenamefont {Yang}(2014)}]{yyp_ijmpa}%
  \BibitemOpen
  \bibfield  {author} {\bibinfo {author} {\bibfnamefont {Y.}~\bibnamefont
  {Yang}},\ }\href {\doibase 10.1142/S0217751X14501942} {\bibfield  {journal}
  {\bibinfo  {journal} {Int. J. Mod. Phys.}\ }\textbf {\bibinfo {volume}
  {A29}},\ \bibinfo {pages} {1450194} (\bibinfo {year} {2014})},\ \Eprint
  {http://arxiv.org/abs/1501.00789} {arXiv:1501.00789 [astro-ph.CO]}
  \BibitemShut {NoStop}%
\bibitem [{\citenamefont {Li}\ \emph {et~al.}(2012)\citenamefont {Li},
  \citenamefont {Erickcek},\ and\ \citenamefont {Law}}]{fangdali}%
  \BibitemOpen
  \bibfield  {author} {\bibinfo {author} {\bibfnamefont {F.}~\bibnamefont
  {Li}}, \bibinfo {author} {\bibfnamefont {A.~L.}\ \bibnamefont {Erickcek}}, \
  and\ \bibinfo {author} {\bibfnamefont {N.~M.}\ \bibnamefont {Law}},\ }\href
  {\doibase 10.1103/PhysRevD.86.043519} {\bibfield  {journal} {\bibinfo
  {journal} {Phys. Rev. D}\ }\textbf {\bibinfo {volume} {86}},\ \bibinfo
  {pages} {043519} (\bibinfo {year} {2012})}\BibitemShut {NoStop}%
\bibitem [{\citenamefont {Clark}\ \emph
  {et~al.}(2016{\natexlab{a}})\citenamefont {Clark}, \citenamefont {Lewis},\
  and\ \citenamefont {Scott}}]{ucmhs_pulsar}%
  \BibitemOpen
  \bibfield  {author} {\bibinfo {author} {\bibfnamefont {H.~A.}\ \bibnamefont
  {Clark}}, \bibinfo {author} {\bibfnamefont {G.~F.}\ \bibnamefont {Lewis}}, \
  and\ \bibinfo {author} {\bibfnamefont {P.}~\bibnamefont {Scott}},\ }\href
  {\doibase 10.1093/mnras/stv2743} {\bibfield  {journal} {\bibinfo  {journal}
  {Mon. Not. Roy. Astron. Soc.}\ }\textbf {\bibinfo {volume} {456}},\ \bibinfo
  {pages} {1394} (\bibinfo {year} {2016}{\natexlab{a}})},\ \Eprint
  {http://arxiv.org/abs/1509.02938} {arXiv:1509.02938 [astro-ph.CO]}
  \BibitemShut {NoStop}%
\bibitem [{\citenamefont {Bird}\ \emph {et~al.}(2011)\citenamefont {Bird},
  \citenamefont {Peiris}, \citenamefont {Viel},\ and\ \citenamefont
  {Verde}}]{lyman}%
  \BibitemOpen
  \bibfield  {author} {\bibinfo {author} {\bibfnamefont {S.}~\bibnamefont
  {Bird}}, \bibinfo {author} {\bibfnamefont {H.~V.}\ \bibnamefont {Peiris}},
  \bibinfo {author} {\bibfnamefont {M.}~\bibnamefont {Viel}}, \ and\ \bibinfo
  {author} {\bibfnamefont {L.}~\bibnamefont {Verde}},\ }\href {\doibase
  10.1111/j.1365-2966.2011.18245.x} {\bibfield  {journal} {\bibinfo  {journal}
  {Monthly Notices of the Royal Astronomical Society}\ }\textbf {\bibinfo
  {volume} {413}},\ \bibinfo {pages} {1717} (\bibinfo {year}
  {2011})}\BibitemShut {NoStop}%
\bibitem [{\citenamefont {Tinker}\ \emph {et~al.}(2012)\citenamefont {Tinker},
  \citenamefont {Sheldon}, \citenamefont {Wechsler}, \citenamefont {Becker},
  \citenamefont {Rozo}, \citenamefont {Zu}, \citenamefont {Weinberg},
  \citenamefont {Zehavi}, \citenamefont {Blanton}, \citenamefont {Busha},\ and\
  \citenamefont {Koester}}]{large}%
  \BibitemOpen
  \bibfield  {author} {\bibinfo {author} {\bibfnamefont {J.~L.}\ \bibnamefont
  {Tinker}}, \bibinfo {author} {\bibfnamefont {E.~S.}\ \bibnamefont {Sheldon}},
  \bibinfo {author} {\bibfnamefont {R.~H.}\ \bibnamefont {Wechsler}}, \bibinfo
  {author} {\bibfnamefont {M.~R.}\ \bibnamefont {Becker}}, \bibinfo {author}
  {\bibfnamefont {E.}~\bibnamefont {Rozo}}, \bibinfo {author} {\bibfnamefont
  {Y.}~\bibnamefont {Zu}}, \bibinfo {author} {\bibfnamefont {D.~H.}\
  \bibnamefont {Weinberg}}, \bibinfo {author} {\bibfnamefont {I.}~\bibnamefont
  {Zehavi}}, \bibinfo {author} {\bibfnamefont {M.~R.}\ \bibnamefont {Blanton}},
  \bibinfo {author} {\bibfnamefont {M.~T.}\ \bibnamefont {Busha}}, \ and\
  \bibinfo {author} {\bibfnamefont {B.~P.}\ \bibnamefont {Koester}},\ }\href
  {http://stacks.iop.org/0004-637X/745/i=1/a=16} {\bibfield  {journal}
  {\bibinfo  {journal} {The Astrophysical Journal}\ }\textbf {\bibinfo {volume}
  {745}},\ \bibinfo {pages} {16} (\bibinfo {year} {2012})}\BibitemShut
  {NoStop}%
\bibitem [{\citenamefont {Josan}\ \emph {et~al.}(2009)\citenamefont {Josan},
  \citenamefont {Green},\ and\ \citenamefont {Malik}}]{Josan:2009}%
  \BibitemOpen
  \bibfield  {author} {\bibinfo {author} {\bibfnamefont {A.~S.}\ \bibnamefont
  {Josan}}, \bibinfo {author} {\bibfnamefont {A.~M.}\ \bibnamefont {Green}}, \
  and\ \bibinfo {author} {\bibfnamefont {K.~A.}\ \bibnamefont {Malik}},\ }\href
  {\doibase 10.1103/PhysRevD.79.103520} {\bibfield  {journal} {\bibinfo
  {journal} {Phys. Rev.}\ }\textbf {\bibinfo {volume} {D79}},\ \bibinfo {pages}
  {103520} (\bibinfo {year} {2009})},\ \Eprint {http://arxiv.org/abs/0903.3184}
  {arXiv:0903.3184 [astro-ph.CO]} \BibitemShut {NoStop}%
\bibitem [{\citenamefont {Gaisser}\ and\ \citenamefont {Honda}(2002)}]{atm_1}%
  \BibitemOpen
  \bibfield  {author} {\bibinfo {author} {\bibfnamefont {T.~K.}\ \bibnamefont
  {Gaisser}}\ and\ \bibinfo {author} {\bibfnamefont {M.}~\bibnamefont
  {Honda}},\ }\href {\doibase 10.1146/annurev.nucl.52.050102.090645} {\bibfield
   {journal} {\bibinfo  {journal} {Ann. Rev. Nucl. Part. Sci.}\ }\textbf
  {\bibinfo {volume} {52}},\ \bibinfo {pages} {153} (\bibinfo {year} {2002})},\
  \Eprint {http://arxiv.org/abs/hep-ph/0203272} {arXiv:hep-ph/0203272 [hep-ph]}
  \BibitemShut {NoStop}%
\bibitem [{\citenamefont {Honda}\ \emph {et~al.}(2007)\citenamefont {Honda},
  \citenamefont {Kajita}, \citenamefont {Kasahara}, \citenamefont
  {Midorikawa},\ and\ \citenamefont {Sanuki}}]{atm_2}%
  \BibitemOpen
  \bibfield  {author} {\bibinfo {author} {\bibfnamefont {M.}~\bibnamefont
  {Honda}}, \bibinfo {author} {\bibfnamefont {T.}~\bibnamefont {Kajita}},
  \bibinfo {author} {\bibfnamefont {K.}~\bibnamefont {Kasahara}}, \bibinfo
  {author} {\bibfnamefont {S.}~\bibnamefont {Midorikawa}}, \ and\ \bibinfo
  {author} {\bibfnamefont {T.}~\bibnamefont {Sanuki}},\ }\href {\doibase
  10.1103/PhysRevD.75.043006} {\bibfield  {journal} {\bibinfo  {journal} {Phys.
  Rev. D}\ }\textbf {\bibinfo {volume} {75}},\ \bibinfo {pages} {043006}
  (\bibinfo {year} {2007})}\BibitemShut {NoStop}%
\bibitem [{\citenamefont {Pasquali}\ and\ \citenamefont
  {Reno}(1999)}]{tau_origi}%
  \BibitemOpen
  \bibfield  {author} {\bibinfo {author} {\bibfnamefont {L.}~\bibnamefont
  {Pasquali}}\ and\ \bibinfo {author} {\bibfnamefont {M.~H.}\ \bibnamefont
  {Reno}},\ }\href {\doibase 10.1103/PhysRevD.59.093003} {\bibfield  {journal}
  {\bibinfo  {journal} {Phys. Rev.}\ }\textbf {\bibinfo {volume} {D59}},\
  \bibinfo {pages} {093003} (\bibinfo {year} {1999})},\ \Eprint
  {http://arxiv.org/abs/hep-ph/9811268} {arXiv:hep-ph/9811268 [hep-ph]}
  \BibitemShut {NoStop}%
\bibitem [{\citenamefont {Ingelman}\ and\ \citenamefont
  {Thunman}(1996)}]{tau_solar}%
  \BibitemOpen
  \bibfield  {author} {\bibinfo {author} {\bibfnamefont {G.}~\bibnamefont
  {Ingelman}}\ and\ \bibinfo {author} {\bibfnamefont {M.}~\bibnamefont
  {Thunman}},\ }\href {\doibase 10.1103/PhysRevD.54.4385} {\bibfield  {journal}
  {\bibinfo  {journal} {Phys. Rev.}\ }\textbf {\bibinfo {volume} {D54}},\
  \bibinfo {pages} {4385} (\bibinfo {year} {1996})},\ \Eprint
  {http://arxiv.org/abs/hep-ph/9604288} {arXiv:hep-ph/9604288 [hep-ph]}
  \BibitemShut {NoStop}%
\bibitem [{\citenamefont {Argüelles}\ \emph {et~al.}(2017)\citenamefont
  {Argüelles}, \citenamefont {de~Wasseige}, \citenamefont {Fedynitch},\ and\
  \citenamefont {Jones}}]{solar_cor_update_1}%
  \BibitemOpen
  \bibfield  {author} {\bibinfo {author} {\bibfnamefont {C.}~\bibnamefont
  {Argüelles}}, \bibinfo {author} {\bibfnamefont {G.}~\bibnamefont
  {de~Wasseige}}, \bibinfo {author} {\bibfnamefont {A.}~\bibnamefont
  {Fedynitch}}, \ and\ \bibinfo {author} {\bibfnamefont {B.}~\bibnamefont
  {Jones}},\ }\href {http://stacks.iop.org/1475-7516/2017/i=07/a=024}
  {\bibfield  {journal} {\bibinfo  {journal} {Journal of Cosmology and
  Astroparticle Physics}\ }\textbf {\bibinfo {volume} {2017}},\ \bibinfo
  {pages} {024} (\bibinfo {year} {2017})}\BibitemShut {NoStop}%
\bibitem [{\citenamefont {Ng}\ \emph {et~al.}(2017)\citenamefont {Ng},
  \citenamefont {Beacom}, \citenamefont {Peter},\ and\ \citenamefont
  {Rott}}]{solar_cor_update_2}%
  \BibitemOpen
  \bibfield  {author} {\bibinfo {author} {\bibfnamefont {K.~C.~Y.}\
  \bibnamefont {Ng}}, \bibinfo {author} {\bibfnamefont {J.~F.}\ \bibnamefont
  {Beacom}}, \bibinfo {author} {\bibfnamefont {A.~H.~G.}\ \bibnamefont
  {Peter}}, \ and\ \bibinfo {author} {\bibfnamefont {C.}~\bibnamefont {Rott}},\
  }\href@noop {} {\  (\bibinfo {year} {2017})},\ \Eprint
  {http://arxiv.org/abs/1703.10280} {arXiv:1703.10280 [astro-ph.HE]}
  \BibitemShut {NoStop}%
\bibitem [{\citenamefont {Edsjö}\ \emph {et~al.}(2017)\citenamefont {Edsjö},
  \citenamefont {Elevant}, \citenamefont {Enberg},\ and\ \citenamefont
  {Niblaeus}}]{solar_cor_update_3}%
  \BibitemOpen
  \bibfield  {author} {\bibinfo {author} {\bibfnamefont {J.}~\bibnamefont
  {Edsjö}}, \bibinfo {author} {\bibfnamefont {J.}~\bibnamefont {Elevant}},
  \bibinfo {author} {\bibfnamefont {R.}~\bibnamefont {Enberg}}, \ and\ \bibinfo
  {author} {\bibfnamefont {C.}~\bibnamefont {Niblaeus}},\ }\href
  {http://stacks.iop.org/1475-7516/2017/i=06/a=033} {\bibfield  {journal}
  {\bibinfo  {journal} {Journal of Cosmology and Astroparticle Physics}\
  }\textbf {\bibinfo {volume} {2017}},\ \bibinfo {pages} {033} (\bibinfo {year}
  {2017})}\BibitemShut {NoStop}%
\bibitem [{\citenamefont {Athar}\ \emph {et~al.}(2005)\citenamefont {Athar},
  \citenamefont {Lee},\ and\ \citenamefont {Lin}}]{tau_galactic}%
  \BibitemOpen
  \bibfield  {author} {\bibinfo {author} {\bibfnamefont {H.}~\bibnamefont
  {Athar}}, \bibinfo {author} {\bibfnamefont {F.-F.}\ \bibnamefont {Lee}}, \
  and\ \bibinfo {author} {\bibfnamefont {G.-L.}\ \bibnamefont {Lin}},\ }\href
  {\doibase 10.1103/PhysRevD.71.103008} {\bibfield  {journal} {\bibinfo
  {journal} {Phys. Rev.}\ }\textbf {\bibinfo {volume} {D71}},\ \bibinfo {pages}
  {103008} (\bibinfo {year} {2005})},\ \Eprint
  {http://arxiv.org/abs/hep-ph/0407183} {arXiv:hep-ph/0407183 [hep-ph]}
  \BibitemShut {NoStop}%
\bibitem [{\citenamefont {Berezinsky}\ \emph {et~al.}(2003)\citenamefont
  {Berezinsky}, \citenamefont {Dokuchaev},\ and\ \citenamefont
  {Eroshenko}}]{ucmhs_2}%
  \BibitemOpen
  \bibfield  {author} {\bibinfo {author} {\bibfnamefont {V.}~\bibnamefont
  {Berezinsky}}, \bibinfo {author} {\bibfnamefont {V.}~\bibnamefont
  {Dokuchaev}}, \ and\ \bibinfo {author} {\bibfnamefont {Y.}~\bibnamefont
  {Eroshenko}},\ }\href {\doibase 10.1103/PhysRevD.68.103003} {\bibfield
  {journal} {\bibinfo  {journal} {Phys. Rev.}\ }\textbf {\bibinfo {volume}
  {D68}},\ \bibinfo {pages} {103003} (\bibinfo {year} {2003})},\ \Eprint
  {http://arxiv.org/abs/astro-ph/0301551} {arXiv:astro-ph/0301551 [astro-ph]}
  \BibitemShut {NoStop}%
\bibitem [{\citenamefont {Gosenca}\ \emph {et~al.}(2017)\citenamefont
  {Gosenca}, \citenamefont {Adamek}, \citenamefont {Byrnes},\ and\
  \citenamefont {Hotchkiss}}]{ucmhs_2017_profile}%
  \BibitemOpen
  \bibfield  {author} {\bibinfo {author} {\bibfnamefont {M.}~\bibnamefont
  {Gosenca}}, \bibinfo {author} {\bibfnamefont {J.}~\bibnamefont {Adamek}},
  \bibinfo {author} {\bibfnamefont {C.~T.}\ \bibnamefont {Byrnes}}, \ and\
  \bibinfo {author} {\bibfnamefont {S.}~\bibnamefont {Hotchkiss}},\ }\href@noop
  {} {\  (\bibinfo {year} {2017})},\ \Eprint {http://arxiv.org/abs/1710.02055}
  {arXiv:1710.02055 [astro-ph.CO]} \BibitemShut {NoStop}%
\bibitem [{\citenamefont {Komatsu}\ \emph {et~al.}(2009)\citenamefont {Komatsu}
  \emph {et~al.}}]{Komatsu:2008hk}%
  \BibitemOpen
  \bibfield  {author} {\bibinfo {author} {\bibfnamefont {E.}~\bibnamefont
  {Komatsu}} \emph {et~al.} (\bibinfo {collaboration} {WMAP}),\ }\href
  {\doibase 10.1088/0067-0049/180/2/330} {\bibfield  {journal} {\bibinfo
  {journal} {Astrophys. J. Suppl.}\ }\textbf {\bibinfo {volume} {180}},\
  \bibinfo {pages} {330} (\bibinfo {year} {2009})},\ \Eprint
  {http://arxiv.org/abs/0803.0547} {arXiv:0803.0547 [astro-ph]} \BibitemShut
  {NoStop}%
\bibitem [{\citenamefont {Clark}\ \emph
  {et~al.}(2016{\natexlab{b}})\citenamefont {Clark}, \citenamefont {Lewis},\
  and\ \citenamefont {Scott}}]{scott_2015}%
  \BibitemOpen
  \bibfield  {author} {\bibinfo {author} {\bibfnamefont {H.~A.}\ \bibnamefont
  {Clark}}, \bibinfo {author} {\bibfnamefont {G.~F.}\ \bibnamefont {Lewis}}, \
  and\ \bibinfo {author} {\bibfnamefont {P.}~\bibnamefont {Scott}},\ }\href
  {\doibase 10.1093/mnras/stw2305, 10.1093/mnras/stv2529} {\bibfield  {journal}
  {\bibinfo  {journal} {Mon. Not. Roy. Astron. Soc.}\ }\textbf {\bibinfo
  {volume} {456}},\ \bibinfo {pages} {1402} (\bibinfo {year}
  {2016}{\natexlab{b}})},\ \bibinfo {note} {[Erratum: Mon. Not. Roy. Astron.
  Soc.464,no.1,955(2017)]},\ \Eprint {http://arxiv.org/abs/1509.02941}
  {arXiv:1509.02941 [astro-ph.CO]} \BibitemShut {NoStop}%
\bibitem [{\citenamefont {Berezinsky}\ \emph
  {et~al.}(2014{\natexlab{a}})\citenamefont {Berezinsky}, \citenamefont
  {Dokuchaev},\ and\ \citenamefont {Eroshenko}}]{ucmhs_inner}%
  \BibitemOpen
  \bibfield  {author} {\bibinfo {author} {\bibfnamefont {V.~S.}\ \bibnamefont
  {Berezinsky}}, \bibinfo {author} {\bibfnamefont {V.~I.}\ \bibnamefont
  {Dokuchaev}}, \ and\ \bibinfo {author} {\bibfnamefont {Y.~N.}\ \bibnamefont
  {Eroshenko}},\ }\href {\doibase 10.3367/UFNe.0184.201401a.0003} {\bibfield
  {journal} {\bibinfo  {journal} {Phys. Usp.}\ }\textbf {\bibinfo {volume}
  {57}},\ \bibinfo {pages} {1} (\bibinfo {year} {2014}{\natexlab{a}})},\
  \bibinfo {note} {[Usp. Fiz. Nauk184,3(2014)]},\ \Eprint
  {http://arxiv.org/abs/1405.2204} {arXiv:1405.2204 [astro-ph.HE]} \BibitemShut
  {NoStop}%
\bibitem [{\citenamefont {Berezinsky}\ \emph
  {et~al.}(2014{\natexlab{b}})\citenamefont {Berezinsky}, \citenamefont
  {Dokuchaev},\ and\ \citenamefont {Eroshenko}}]{profile}%
  \BibitemOpen
  \bibfield  {author} {\bibinfo {author} {\bibfnamefont {V.~S.}\ \bibnamefont
  {Berezinsky}}, \bibinfo {author} {\bibfnamefont {V.~I.}\ \bibnamefont
  {Dokuchaev}}, \ and\ \bibinfo {author} {\bibfnamefont {Y.~N.}\ \bibnamefont
  {Eroshenko}},\ }\href {\doibase 10.3367/UFNe.0184.201401a.0003} {\bibfield
  {journal} {\bibinfo  {journal} {Phys. Usp.}\ }\textbf {\bibinfo {volume}
  {57}},\ \bibinfo {pages} {1} (\bibinfo {year} {2014}{\natexlab{b}})},\
  \bibinfo {note} {[Usp. Fiz. Nauk184,3(2014)]},\ \Eprint
  {http://arxiv.org/abs/1405.2204} {arXiv:1405.2204 [astro-ph.HE]} \BibitemShut
  {NoStop}%
\bibitem [{\citenamefont {Bergstrom}\ \emph {et~al.}(1997)\citenamefont
  {Bergstrom}, \citenamefont {Edsjo},\ and\ \citenamefont
  {Gondolo}}]{Bergstrom:1996kp}%
  \BibitemOpen
  \bibfield  {author} {\bibinfo {author} {\bibfnamefont {L.}~\bibnamefont
  {Bergstrom}}, \bibinfo {author} {\bibfnamefont {J.}~\bibnamefont {Edsjo}}, \
  and\ \bibinfo {author} {\bibfnamefont {P.}~\bibnamefont {Gondolo}},\ }\href
  {\doibase 10.1103/PhysRevD.55.1765} {\bibfield  {journal} {\bibinfo
  {journal} {Phys. Rev.}\ }\textbf {\bibinfo {volume} {D55}},\ \bibinfo {pages}
  {1765} (\bibinfo {year} {1997})},\ \Eprint
  {http://arxiv.org/abs/hep-ph/9607237} {arXiv:hep-ph/9607237 [hep-ph]}
  \BibitemShut {NoStop}%
\bibitem [{\citenamefont {Arcadi}\ \emph {et~al.}(2017)\citenamefont {Arcadi},
  \citenamefont {Dutra}, \citenamefont {Ghosh}, \citenamefont {Lindner},
  \citenamefont {Mambrini}, \citenamefont {Pierre}, \citenamefont {Profumo},\
  and\ \citenamefont {Queiroz}}]{Arcadi:2017kky}%
  \BibitemOpen
  \bibfield  {author} {\bibinfo {author} {\bibfnamefont {G.}~\bibnamefont
  {Arcadi}}, \bibinfo {author} {\bibfnamefont {M.}~\bibnamefont {Dutra}},
  \bibinfo {author} {\bibfnamefont {P.}~\bibnamefont {Ghosh}}, \bibinfo
  {author} {\bibfnamefont {M.}~\bibnamefont {Lindner}}, \bibinfo {author}
  {\bibfnamefont {Y.}~\bibnamefont {Mambrini}}, \bibinfo {author}
  {\bibfnamefont {M.}~\bibnamefont {Pierre}}, \bibinfo {author} {\bibfnamefont
  {S.}~\bibnamefont {Profumo}}, \ and\ \bibinfo {author} {\bibfnamefont
  {F.~S.}\ \bibnamefont {Queiroz}},\ }\href@noop {} {\  (\bibinfo {year}
  {2017})},\ \Eprint {http://arxiv.org/abs/1703.07364} {arXiv:1703.07364
  [hep-ph]} \BibitemShut {NoStop}%
\bibitem [{\citenamefont {Mandal}\ \emph {et~al.}(2010)\citenamefont {Mandal},
  \citenamefont {Buckley}, \citenamefont {Freese}, \citenamefont {Spolyar},\
  and\ \citenamefont {Murayama}}]{tau_detec}%
  \BibitemOpen
  \bibfield  {author} {\bibinfo {author} {\bibfnamefont {S.~K.}\ \bibnamefont
  {Mandal}}, \bibinfo {author} {\bibfnamefont {M.~R.}\ \bibnamefont {Buckley}},
  \bibinfo {author} {\bibfnamefont {K.}~\bibnamefont {Freese}}, \bibinfo
  {author} {\bibfnamefont {D.}~\bibnamefont {Spolyar}}, \ and\ \bibinfo
  {author} {\bibfnamefont {H.}~\bibnamefont {Murayama}},\ }\href {\doibase
  10.1103/PhysRevD.81.043508} {\bibfield  {journal} {\bibinfo  {journal} {Phys.
  Rev. D}\ }\textbf {\bibinfo {volume} {81}},\ \bibinfo {pages} {043508}
  (\bibinfo {year} {2010})}\BibitemShut {NoStop}%
\bibitem [{\citenamefont {{Wiebusch}}\ and\ \citenamefont {{for the IceCube
  Collaboration}}(2009)}]{2009arXiv0907.2263W}%
  \BibitemOpen
  \bibfield  {author} {\bibinfo {author} {\bibfnamefont {C.}~\bibnamefont
  {{Wiebusch}}}\ and\ \bibinfo {author} {\bibnamefont {{for the IceCube
  Collaboration}}},\ }\href@noop {} {\bibfield  {journal} {\bibinfo  {journal}
  {ArXiv e-prints}\ } (\bibinfo {year} {2009})},\ \Eprint
  {http://arxiv.org/abs/0907.2263} {arXiv:0907.2263 [astro-ph.IM]} \BibitemShut
  {NoStop}%
\bibitem [{\citenamefont {Coyle}\ and\ \citenamefont
  {James}(2017)}]{Coyle:2017pnl}%
  \BibitemOpen
  \bibfield  {author} {\bibinfo {author} {\bibfnamefont {P.}~\bibnamefont
  {Coyle}}\ and\ \bibinfo {author} {\bibfnamefont {C.~W.}\ \bibnamefont
  {James}} (\bibinfo {collaboration} {ANTARES})\ }(\bibinfo {year} {2017})\
  \Eprint {http://arxiv.org/abs/1701.02144} {arXiv:1701.02144 [astro-ph.HE]}
  \BibitemShut {NoStop}%
\bibitem [{\citenamefont {Jeong}\ and\ \citenamefont {Reno}(2010)}]{sigma_cc}%
  \BibitemOpen
  \bibfield  {author} {\bibinfo {author} {\bibfnamefont {Y.~S.}\ \bibnamefont
  {Jeong}}\ and\ \bibinfo {author} {\bibfnamefont {M.~H.}\ \bibnamefont
  {Reno}},\ }\href {\doibase 10.1103/PhysRevD.82.033010} {\bibfield  {journal}
  {\bibinfo  {journal} {Phys. Rev.}\ }\textbf {\bibinfo {volume} {D82}},\
  \bibinfo {pages} {033010} (\bibinfo {year} {2010})},\ \Eprint
  {http://arxiv.org/abs/1007.1966} {arXiv:1007.1966 [hep-ph]} \BibitemShut
  {NoStop}%
\bibitem [{\citenamefont {Zhu}(2006)}]{Zhu:2008yh}%
  \BibitemOpen
  \bibfield  {author} {\bibinfo {author} {\bibfnamefont {Y.-S.}\ \bibnamefont
  {Zhu}},\ }\href@noop {} {\bibfield  {journal} {\bibinfo  {journal} {HEPNP}\
  }\textbf {\bibinfo {volume} {30}},\ \bibinfo {pages} {331} (\bibinfo {year}
  {2006})},\ \Eprint {http://arxiv.org/abs/0812.2708} {arXiv:0812.2708
  [physics.data-an]} \BibitemShut {NoStop}%
\end{thebibliography}
%

\end{document}